%% file: main.tex
\documentclass[letterpaper, 10pt, conference]{ieeeconf}
\IEEEoverridecommandlockouts
\usepackage{import}

\usepackage{algorithm,algpseudocode}
\usepackage{verbatim}
\usepackage{hyperref}

\let\labelindent\relax
\usepackage{enumitem}
\usepackage{algpseudocode}
\usepackage{amsmath,amsfonts,amssymb}     
\usepackage{graphicx}
\usepackage{mathabx}
\usepackage{csquotes}
\usepackage{bm}
\usepackage{subcaption}
\usepackage{comment}
\usepackage{xcolor}
\usepackage{tabu}
\usepackage{bbm}
\usepackage{wrapfig}
\usepackage{steinmetz}
\graphicspath{{./figure/}}

\newtheorem{theorem}{Theorem}
\newtheorem{lemma}{Lemma}

\newtheorem{definition}{Definition}
\newtheorem{assump}{Assumption}

\newtheorem{proposition}{Proposition}
\newtheorem{corollary}{Corollary}
\newtheorem{remark}{Remark}

\newcommand{\eps}{\varepsilon}

\newcommand{\C}{\mathbf{C}}
\newcommand{\Z}{\mathbb{Z}}
\newcommand{\I}{\mathcal{I}}
\newcommand{\N}{\mathbb{N}}
\newcommand{\R}{\mathbb{R}}

\newcommand{\Linf}[1]{\left\|{#1}\right\|_{\infty}}
\newcommand{\linf}[1]{\|{#1}\|_{\infty}}

\newcommand{\black}[1]{\textcolor{black}{#1}}

\def\X{{\mathbf X}}
\def\x{{\mathbf x}}

\def\g{{\boldsymbol \gamma}}

\def\A{{\mathbf A}}
\def\B{{\mathbf B}}
\def\C{{\mathbf C}}
\def\D{{\mathbf D}}

\def\C{{\mathbf C}}

\def\e{{\mathbf e}}
\def\H{{\mathbf H}}
\def\I{{\mathbf I}}
\def\J{{\mathbf J}}
\def\K{{\mathbf K}}
\def\L{{\mathbf L}}
\def\M{{\mathbf M}}
\def\N{{\mathbf N}}
\def\P{{\mathbf P}}
\def\Q{{\mathbf Q}}
\def\S{{\mathbf S}}
\def\U{{\mathbf U}}
\def\V{{\mathbf V}}
\def\Y{{\mathbf Y}}
\def\Z{{\mathbf Z}}
\def\S{{\mathbf S}}
\def\W{{\mathbf W}}

\def\bxi{{\mathbf \xi}}
\def\zero{{\mathbf 0}}
\def\1{{\mathbbm 1}}

\def\mV{{\mathcal V}}

\def\mC{{\mathcal C}}
\def\mE{{\mathcal E}}

\def\mG{{\mathcal G}}

\def\mP{{\mathcal P}}

\def\mS{{\mathcal S}}
\def\mT{{\mathcal T}}
\def\wx{ {\widetilde \x}}

\def\hS{ {\widehat \S}}
\def\hL{ {\widehat \L}}

\def\mbS{{\mathbb S}}

\def\tS{\mathbf {\tilde{S}}}
\def\tL{\mathbf {\tilde{L}}}

\def\dh{d_{hop}}

\def\<{{\langle}}
\def\>{{\rangle}}

\def\O{{\Omega}}

\usepackage{titlesec}
\titlespacing\section{-5pt}{12pt plus 4pt minus 2pt}{0pt plus 2pt minus 2pt}
\titlespacing\subsection{0pt}{12pt plus 4pt minus 2pt}{0pt plus 2pt minus 2pt}
\titlespacing\subsubsection{0pt}{12pt plus 4pt minus 2pt}{0pt plus 2pt minus 2pt}

\title{Topology Learning of Linear Dynamical Systems with Latent Nodes using Matrix Decomposition}

\author{Mishfad Shaikh Veedu$^{*,1}$, \emph{Student Member, IEEE}, Harish Doddi$^{*,2}$, and Murti V. Salapaka$^{1}$, \emph{Fellow, IEEE}\\
	\thanks{ $^{1}$Mishfad S. Veedu and Murti V. Salapaka are with Department of Electrical and Computer Engineering, University of Minnesota, Minneapolis, USA,
		{\tt\small veedu002@umn.edu, murtis@umn.edu}}
	\thanks{$^{2}$Harish Doddi is with Department of Mechanical Engineering, University of Minnesota, Minneapolis, USA,
		{\tt\small doddi003@umn.edu}}%
	\thanks{$^{*}$Mishfad S. Veedu and Harish Doddi contributed equally to this work}}
\begin{document}
\maketitle\relax
{	
\begin{abstract}
\black{In this article, we present a novel approach to reconstruct the topology of networked linear dynamical systems with latent nodes. The network is allowed to have directed loops and bi-directed edges. The main approach relies on the unique decomposition of the inverse of power spectral density matrix (IPSDM) obtained from observed nodes as a sum of sparse and low-rank matrices. We provide conditions and methods for decomposing the IPSDM of the observed nodes into sparse and low-rank components. The sparse component yields the moral graph associated with the observed nodes, and the low-rank component retrieves parents, children and spouses (the Markov Blanket) of the hidden nodes. The article provides necessary and sufficient conditions for the unique decomposition of a given skew symmetric matrix into sum of a sparse skew symmetric and a low-rank skew symmetric matrices. It is shown that for a large class of systems, the unique decomposition of imaginary part of the IPSDM of observed nodes, a skew symmetric matrix, into the sparse and the low-rank components is sufficient to identify the moral graph of the observed nodes as well as the Markov Blanket of latent nodes. For a large class of systems, all spurious links in the moral graph formed by the observed nodes can be identified. Assuming conditions on hidden nodes required for identifiability, links between the hidden and observed nodes can be reconstructed, resulting in the retrieval of the exact topology of the network from the availability of IPSDM.} Moreover, for finite number of data samples, we provide concentration bounds on the entry-wise distance between the true IPSDM and the estimated IPSDM.


\end{abstract}	
}
\section{Introduction}
\label{sec:introduction}
Networks provide convenient representation of large scale complex systems utilized in diverse areas such as power grids, biology, finance, and neuroscience. Reconstructing the underlying network topology or the influence structure of the interaction from measurements is useful in predicting, and steering the behavior of the system towards a desired state. Learning the unknown interaction structure of a network of agents from time-series measurements can be categorized as being active \cite{He_JMLR_08} or passive \cite{Buntine_TKDD_96}. Active techniques \nocite{kekatos} require intervention in the normal operation of the system by injection of external signals and/or altering the network structure, by removing or adding agents to the network. Many critical systems such as the power grid, the financial markets, and the meteorological system either do not allow for active interventions or it is not possible to affect the system. On the contrary, passive techniques infer topology from time-series measurements, without affecting the underlying grid network. Here, in practice, observing time-series measurement at every node is not plausible, wherein it becomes important to learn the topology of the network when only a subset of the nodes are observed.

Learning the topology of a network from time-series data is an active area of research, with considerable emphasis from the machine learning and probabilistic graphical model communities (see \cite{He_JMLR_08}\nocite{Buntine_TKDD_96}\nocite{Elena}\nocite{Yanning_top_id}-\cite{jordan}). However, here, most works assume that the nodes are random variables which fail to capture the dynamics of the interaction and thus are improper for the applications with dynamic dependencies that are common, for example, in the power grid \cite{wood2013power} and its application to smart grid networks \cite{patel2017distributed}, climate science \cite{climate} and finance \cite{thermal}. {\color{black}Moreover, in many scenarios, topology identification is the first step in system identification\cite{VANDENHOF20132994,Ramaswamy}.
}

Filtering based topology reconstruction has gained considerable attention recently for unveiling the topology of dynamically related agents. In \cite{materassi_tac12}, the moral graph of a directed network is reconstructed using the magnitude response of multivariate Wiener filters. \cite{TALUKDAR_physics} provided a method that showed that the spurious links present in the moral graph can be removed by checking the phase response of the Wiener filter between the links. \cite{materassi_tac12} and \cite{materassi_GC} provided algorithms for exact network reconstruction where all parent-child relations are uncovered but the results are restricted using Granger causality to systems with strictly causal dynamical dependencies. \nocite{Talukdar_consensus_net} The aforementioned works, \cite{materassi_tac12}-\nocite{TALUKDAR_physics}\cite{materassi_GC}, assumed full network observability. Several works, \cite{Talukdar_ACC18}\nocite{materassi_blindlearning}\nocite{hidden_polytree}-\nocite{materassi_polytree}\nocite{Salapaka_polytrees}\cite{Salapaka_Id_unobserved}, have studied topology identification in the presence of hidden nodes, but, restricted to radial tree topologies--characterized by undirected tree topology in \cite{Talukdar_ACC18},
for polyforest networks in \cite{materassi_blindlearning}, and polytree networks in \cite{hidden_polytree}-\nocite{materassi_polytree}\cite{Salapaka_polytrees}.
{\nocite{Salapaka_signal_selection,Model_id_CS}}
Network reconstruction with corrupted data streams was explored in \cite{subramanian2017network}-\nocite{subramanian2018inferring}\cite{subramanian2019corruption}\nocite{veedu2020spatial_corre}.

In \cite{zorzi_AR_hidden}-\nocite{AR1,AR2}\nocite{Alpago_ContLett_2018}\nocite{Ciccone_TAC_2019}\nocite{Ciccone_TAC_2020}\cite{AR3}, the authors considered the problem of estimating conditional dependency structure of autoregressive (AR) Gaussian stochastic processes in the presence of latent nodes, {\color{black}with an emphasis on finiteness of the time-series available. Here, the problem is formulated in terms of} sparse plus low-rank decomposition of the inverse of the power spectral density matrix \black{(IPSDM)}. {\color{black}The articles \cite{zorzi_AR_hidden}-\nocite{AR1}\nocite{AR2}\cite{AR3} provided interesting optimization frameworks and theoretical guarantees to identify the conditional dependency from estimated IPSDM. 
Here, the graphical representations of conditional dependencies reconstructed from the IPSDM retrieves the moral graph \cite{materassi_tac12}. As shown in \cite{materassi_tac12,Talukdar_ACC17}, moral graphs can admit many spurious edges.

 In this article, we approach the problem of reconstructing the topology
 in the presence of latent nodes. Similar to \cite{zorzi_AR_hidden}-\nocite{AR1,AR2}\nocite{Alpago_ContLett_2018}\nocite{Ciccone_TAC_2019}\nocite{Ciccone_TAC_2020}\cite{AR3}  a perspective of sparse plus low-rank decomposition of IPSDM associated with the observed nodes is taken.
 
Extending the results in \cite{CSPW_siam_2011}, this article establishes conditions for a skew symmetric to admit a unique decomposition as a sparse and low-rank matrix. Towards  decomposing a skew symmetric matrix, this article provides a characterization for tangent manifolds of skew symmetric matrices with a fixed rank and with a given sparsity pattern. {\color{black}Furthermore, the article provides additional theoretical insights into an empirical procedure presented in \cite{CSPW_siam_2011}, which provides a sufficient condition that can be tractably assessed for a unique decomposition. Though not emphasized in the article, the methodology developed can be used to realize similar results for unique decomposition of Hermitian matrices as well.}
 
{\color{black} Based on the extensions of the results in \cite{CSPW_siam_2011} established here, it is possible to obtain a decomposition of the IPSDM of the observed nodes, where the sparse part can be leveraged to realize the moral graph formed by the observed nodes. We establish identifiability conditions under which the low-rank component of the observed nodes' IPSDM yields the parents, children and spouses (all the the two hop neighbors that form the Markov Blanket) of the latent (hidden) nodes. The IPSDM, being complex and frequency dependent, has real and imaginary parts. We further emphasize the imaginary part of the IPSDM in the article as it has considerable structure applicable for a large class of problems. Here too the sparse component of the imaginary part of the IPSDM matrix, which is skew symmetric, is shown to yield the moral graph of networks governed by a linear dynamical model (defined later) that encompass a wide class of systems. The above approach can be employed toward the  retrieval of the moral graph of networks of  AR models and is applicable to  networked systems addressed in \cite{zorzi_AR_hidden} albeit, here we do not emphasize the finite data aspects.

 The moral graph relations can admit many spurious connections not present in the original topology. We demonstrate that the rank-sparsity patterns induced by the network topology on the imaginary part has properties that can be exploited toward the {\it exact} reconstruction of network topology. Here, under assumptions applicable for a large class of systems, all spurious links in the moral graph formed by observed nodes can be identified. Moreover, assuming conditions on hidden (latent) nodes, which follow from identifiability conditions, links between the hidden and observed nodes can be reconstructed resulting in the retrieval of the exact topology of the network.
 
 This article also serves as an important bridge between the works presented in \cite{CSPW_siam_2011} and the works related to network structure reconstruction \cite{materassi_tac12}-\cite{Model_id_CS}. Furthermore, results here are applicable to many classes of directed graphs without self loops, not restricted to directed acyclic graphs or bi-directed graphs, unlike \cite{TALUKDAR_physics,Talukdar_ACC18}.
 }

}

We summarize below the major contributions of the article.
	\begin{itemize}
{\color{black}\item Provides non-trivial generalizations of results of \cite{CSPW_siam_2011} to skew symmetric matrices, with exact characterization of tangent manifolds of skew symmetric matrices with a given rank and skew symmetric sparse matrices. We also provide a sufficient condition that enables a practical way to select penalty factor for the convex optimization formulation that yields the unique, sparse plus low-rank matrix decomposition, restricted to skew symmetric matrices. This contribution is applicable to general skew symmetric matrix independent of its application to topology identification. 
\item Reconstructs the moral graph of observed nodes and the Markov Blanket of latent nodes from the matrix decomposition of IPSDM; accounting for network identifiability issues associated with the latent nodes.
\item Establishes that the decomposition of imaginary part of IPSDM is sufficient to recover moral graph of observed nodes and the Markov Blanket of the latent nodes. Conditions and methods for unique decomposition as a sum of low rank and sparse matrices of the imaginary part are provided.
\item For large class of systems, the exact topology of the entire network is reconstructed by the decomposition of imaginary part of the IPSDM.
\item For the more practical scenarios where we have access only to finite samples of time-series at each node, we provide a concentration bound for estimation error of IPSDM.}
\end{itemize}

\black{Organization of the article: Section \ref{sec:matrix_decomp} presents the unique decomposition of a skew symmetric matrix into its sparse and low-rank components. Section \ref{sec:LD_system} discusses linear dynamical systems and its graphical representation. Section \ref{sec:moral_graph} addresses moral graph reconstruction, identifiability of latent nodes, and Markov Blanket reconstruction of the latent nodes. Exact topology reconstruction is studied in Section \ref{sec:top_rec}. Section \ref{sec:finite_TS} discusses IPSDM estimation from finite time series. Section \ref{sec:simulation} provides simulation results and Section \ref{sec:conclusion} concludes the article.
} 

\emph{Notations:} Bold capital letters denote matrices and bold small letters denote vectors. $\mbS^n$ denotes the set of all $n \times n$ skew symmetric matrices with real entries. For a matrix $\M$, $[\M]_{ij}$, $\M_{ij}$, or $\M(i,j)$ denote the $(i,j)$-th element of $\M$, $ \| \M \|_1$ is defined as $\sum_{i,j} |\M_{ij}|$, $ \| \M \|_{\infty}$ is defined as $\max_{ij} |\M_{ij}|$, and $\|\M\|_0$ denotes the number of non-zero entries in $\M$. $ \| \M \|_{*}$ denotes the nuclear norm, which is the sum of singular values of $\M$, and $ \| \M \|_2$ denotes spectral-norm, which is defined as the largest singular value. $support(\M)$ is defined as $\{(i,j): \M_{ij}\neq 0\}$. $\Im\{ \M \}$ denotes imaginary part of $\M.$ For a vector, $ \|\x\|_2$ denotes euclidean-norm, defined as $ \sqrt{\sum_i x^2_i}$. $\sigma_i(\M)$ denotes $i^\text{th}$ largest singular value of $\M$. For time-series $(\widetilde{x}_i(t))_{t\in \mathbb{Z}}$, $\x(z) = \mathcal{Z}[{\wx(k)}]$ denotes bilateral z-transform of $\widetilde{\x}$. For a set $\mS$, $|\mS|$ denotes cardinality of the set. {\color{black} We use $j=\sqrt{-1}$. For a transfer function $h(z), z\in \mathbb{C}$, $h=0$ means $h$ is identically zero, i.e., $h(z)=0$ for every $|z|=1$. $h\neq0$ means $h$ is not identically zero. Almost always or almost surely is defined for a probability measure that is absolutely continuous with respect to the Lebesgue measure (e.g. any continuous distribution).}

\section{Sparse plus Low-Rank Matrix Decomposition for Skew Symmetric Matrices}
\label{sec:matrix_decomp}

In this section, we discuss the following problem: suppose we are given a real skew symmetric matrix $\C \in \mathbb{S}^{n} $ that is obtained by adding a sparse matrix $\tS \in \mathbb{S}^{n}$ and a low-rank matrix $\tL \in \mathbb{S}^{n}$; when can we decompose the matrix and retrieve the component matrices? The material presented in this section provides the needed preliminaries and extensions of results from \cite{CSPW_siam_2011}, which do not incorporate constraints of skew symmetry. {\color{black} We remark that the results discussed in this section for the space of skew symmetric matrices can be extended to the space of complex matrices. However, we do not discuss it here.}
\vspace{-4pt}
\subsection{Optimization for Sparse plus Low-rank Decomposition}
\label{subsec:opti_SL}

Consider the following optimization problem.
{\footnotesize
\begin{eqnarray}
\label{eq:convex_opti0}
(\hS_{\g},\hL_{\g})&=& \arg \min_{\S,\L} \gamma \|\S\|_0 + rank(\L)\\
\nonumber
\hspace{1.5cm}&& \text{subject to }
\nonumber
\S+\L=\C,\\
&& \nonumber
\hspace{1.53cm} \S+\S^T=\mathbf{0}, \ \L+\L^T=\mathbf{0},
\end{eqnarray}}where $\gamma$ is a fixed penalty, selected a priori.
\eqref{eq:convex_opti0} is a combinatorial optimization problem and is NP-hard \cite{integer_programming}. $\ell_1$ norm is often employed as a surrogate for $\ell_0$ norm \cite{geethu2019thesis}, with nuclear norm being a proxy for rank \cite{fazel_thesis}. Thus a more tractable convex relaxation associated with \eqref{eq:convex_opti0} is:
\vspace{-10pt}

{\footnotesize
\begin{eqnarray}
\label{eq:convex_opti}
(\widehat{\S}_{\g},\widehat{\L}_{\g})&=& \arg \min_{\S,\L} \g \|\S\|_1 + \|\L\|_{*}\\
\nonumber
&& \text{subject to } \S+\L=\C,\\
\nonumber
&& \hspace{1.53cm} \S+\S^T=\mathbf{0}, \ \L+\L^T=\mathbf{0}.
\end{eqnarray}}
\vspace{-15pt}
\begin{remark}
Given $\C=-\C^T$, imposing the constraint $\S+\S^T=\zero$ renders $\L+\L^T$ superfluous. 
\end{remark}

In this article, the convex formulation in \eqref{eq:convex_opti} is applied for retrieving the sparse and low-rank components from the given $\C$.
\vspace{-10pt}
\subsection{Affine Varieties and Tangent Spaces} 
\label{subsec:alg_variety_TS}
In the seminal work \cite{CSPW_siam_2011}, the convex optimization problem \eqref{eq:convex_opti} without the constraints $\S^T=-\S$ and $\L^T=-\L$ is considered, which provided sufficient conditions to retrieve $\tS$ and $\tL$ exactly. The results in \cite{CSPW_siam_2011} established results for general square matrices with a real field. In this section, we extend the results to skew symmetric matrices in real field, $\mbS^n$. In order to address the decomposition, we consider the sparse matrix sets as an affine variety and low-rank matrix sets as a manifold. We characterize the necessary and sufficient conditions required for the unique decomposition in terms of the tangent space to the affine variety of support constrained skew symmetric matrices at $\tS$--the original sparse matrix--and the tangent space to the manifold of rank constrained skew symmetric matrices at $\tL$--the original low-rank matrix. Note that an affine variety is defined as the zero set of a system of polynomial equations \cite{IVA}.
\begin{remark}
\label{rem:eig_vals_skew_sym}
The skew symmetric matrices with real entries have the property that all the non-zero eigenvalues are pure imaginary and they exist in conjugate pairs. Therefore, the rank $r$ of every skew symmetric matrix must be \emph{even}, and the multiplicity of the singular values must be a positive multiple of two. 
\end{remark}

Next, we provide definitions of tangent spaces, specific to skew symmetric matrices in real field, as our major focus in this article is on imaginary part of Hermitian matrices. The affine variety of skew symmetric matrices constrained by support size $m$ is defined as: 
\begin{eqnarray}
\mathcal{S}(m):=\{\M \in \mathbb{\mbS}^{n} : |support(\M)|\leq m\}.
\end{eqnarray}
Notice that $\mathcal{S}(m)$ is defined over the space of all skew symmetric matrices, $\mathbb{S}^n$, i.e., the set of all matrices $\M \in \mathbb{R}^{n \times n}$ with $\M^T=-\M$. We establish the following result on tangent spaces of sparse real skew symmetric matrices.
\begin{lemma}
\label{lem:TS_sparse}
For any skew symmetric matrix $\M \in \mathbb{S}^n$, the tangent space $\O(\M)$ with respect to $\mathcal{S}(|support(\M)|)$ at $\M$ is:
\begin{eqnarray}
\O(\M):=\{\N \in \mathbb{S}^{n}: support(\N)  \subseteq support(\M)\}.
\end{eqnarray}
\end{lemma}
\hspace{-0.65cm}\begin{proof}
See Appendix \ref{App:TS_sparse}.
\end{proof}
The dimension of this tangent space is $support(\M)/2$ owing to the skew symmetric property.

We define set of skew symmetric matrices of rank $r$ as:
\begin{eqnarray}
\mathcal{R}(r):=\{\M\in \mathbb{S}^{n} : rank(\M) = r\}.
\end{eqnarray}
It is shown in \cite{kozhasov2020minimality} that $\mathcal{R}(r)$ is a differential manifold, whose dimension is $nr-\frac{r^2-r}{2}$. 
\begin{lemma}
\label{lem:TS_rank}
For any skew symmetric matrix $\M \in \mathbb{S}^n$, the tangent space $T(\M)$ with respect to $\mathcal{R}(rank(\M))$ at $\M$ is:
{\footnotesize
\begin{eqnarray}
\label{eq:def_TS_LR}
T(\M):=\{\U\X^T - \X \U^T : \X\in \mathbb{R}^{n\times r} \},
\end{eqnarray}} 
where $\M=\U\D\V^T$ is the compact singular value decomposition (CSVD) of $\M$.
\end{lemma}
\hspace{-0.65cm}\begin{proof}
See Appendix \ref{app:TS_rank}.
\end{proof}
The dimension of $T(\M)$ is $nr-\frac{r^2-r}{2}$. 

The following lemma is obtained based on Remark \ref{rem:eig_vals_skew_sym}.
\begin{lemma}
\label{lem:proj_UV}
Let $\M\in \mathbb{S}^n$ be a skew symmetric matrix and let $\M=\U\D\V^T$ be CSVD of $\M$, $\D:=diag(\sigma_1,\dots,\sigma_r)$ with $\sigma_1\geq\sigma_2\geq \dots\geq\sigma_r $, where $r$ is the rank of $\M$. Then, the projection matrices $\U\U^T$ and $\V\V^T$ of the given skew symmetric matrix $\M$ are equal. 
\end{lemma}
\hspace{-0.65cm}\begin{proof}
See supplementary material\black{, Appendix A} or \cite{doddi2019topology}. 
\end{proof}

Suppose we have prior information about $\Omega(\tS)$ and $T(\tL)$, in addition to being given $\C=\tS+\tL$. Then, it can be shown that a necessary and sufficient condition for unique identifiability of $(\tS,\tL)$ in terms of the tangent spaces is

{\footnotesize
\begin{equation}
\label{eq:transverse_intersection}
\Omega(\tS) \cap T(\tL)=\{\mathbf{0}\},
 \vspace{-5pt}	
\end{equation}}
i.e., the tangent spaces $\Omega(\tS)$ and $T(\tL)$ has {\color{black}trivial intersection}. In other words, if the tangent spaces intersect only at origin, then we can retrieve the component matrices $\tS$ and $\tL$, if we have access to $\Omega(\tS)$ and $T(\tL)$. Given exact characterization of tangent spaces for real skew symmetric matrices in Lemma \ref{lem:TS_sparse} and Lemma \ref{lem:TS_rank}, it is possible to test for the necessary and sufficient {\color{black}trivial intersection} condition for a given matrix $\M.$ Now, analogous to development in \cite{CSPW_siam_2011}, we obtain the sparse and low-rank decomposition using convex optimization.
	
	%
\vspace{-5pt}
\subsection{Sparse plus low-rank Decomposition using Optimization}
In general, it is not possible to recover the original sparse and low-rank matrices by solving \eqref{eq:convex_opti}. To begin with, the solution of the optimization problem depends intricately on the penalty factor $\g$. In Proposition \ref{prop:diff_t}, we prove that for $\g$ close to zero the optimal solution $(\widehat{\S}_{\g},\widehat{\L}_{\g})$ returned by \eqref{eq:convex_opti} is $(\C,\mathbf{0})$, whereas for $\g$ sufficiently large $(\widehat{\S}_{\g},\widehat{\L}_{\g})=(\mathbf{0},\C)$. 
	
Another issue in decomposing the given matrix $\C$ is if either $\tS$ is low-rank, or $\tL$ is sparse. For example, suppose that the low-rank matrix $\tL$ is such that $\tL_{11}\neq0$ with every other entry zero, and $\tS$ be any sparse matrix with $\tS_{11}\neq 0$. Then, the optimization may return $(\C,\mathbf{0})$ or $(\mathbf{0},\C)$ as the solution depends on the rank of $\tS$. Another example where the unique decomposition is not possible is when $\tS$ has support restricted to the first column and the first column of $\tL$ negates all the entries of $\S$. Then, a reasonable solution is $(\mathbf{0},\C)$. 
	
Next, we characterize the optimal regions of \eqref{eq:convex_opti} and provide sufficient conditions under which it obtain the unique decomposition, i.e., returns the true sparse and low-rank matrices.
The following proposition provides a sufficient condition for \eqref{eq:convex_opti} to return the optimum solution $(\hS_{\g},\hL_{\g})=(\tS,\tL)$.

\begin{proposition}
\label{prop:lagrangian_suff}
Suppose that $\C=\tS+\tL$, $\C\neq \zero$ where $\tS,\tL \in \mbS^n$ is given. Then, $(\tS,\tL)$ is the unique optimizer of \eqref{eq:convex_opti} if the following conditions are satisfied:

{\footnotesize
\begin{enumerate}
    \item $\Omega(\tS{}) \cap T(\tL)=\{\mathbf{0}\}$.
    \item There exist duals $\Q_1, \Q_2 \in \mathbb{R}^{n \times n}$ such that 
    \begin{enumerate}[label=(\alph*)]
\item $P_{\Omega(\tS)}(\Q_1-\Q_2-\Q_2^T)=\gamma sign(\tS)$,
\item $P_{T(\tL)}(\Q_1)=\U\V^T$,
\item $\|P_{\Omega(\tS)^C}(\Q_1-\Q_2-\Q_2^T )\|_{\infty} <  \gamma$,
\item $\|P_{T^{\perp}}(\Q_1)\|_{2}< 1$,
\end{enumerate}
\end{enumerate}}
{\color{black}\noindent where $P_{\Omega(\tS)}(\M)$ is obtained by setting entries of $\M$ outside the support of $\tS$ to zero and projecting it to the space of skew symmetric matrices, and $P_{T(\tL)}(\M ):=P_{\U}\M+\M P_{\U}-P_{\U}\M P_{\U}$; $P_{\U}=\U\U^T$, $\tL=\U\Sigma\V^T$.}
\end{proposition}
\hspace{-0.65cm}\begin{proof}See supplementary material\black{, Appendix B} or \cite{doddi2019topology}. 
\end{proof}
\vspace{-10pt}

\subsection{Sufficient Conditions to Retrieve $\tS$ and $\tL$ }
\label{subsec:suff_cond}
Here, we provide some sufficient conditions that guarantee the existence of the duals $\Q_1$ and $\Q_2$ discussed in Proposition \ref{prop:lagrangian_suff}.
The definitions $\mu(\tS):=\max_{\N \in \Omega(\tS):\|\N\|_{\infty}\leq 1}  \|\N\|_2$ and $\xi(\tL):=\max_{\N \in T(\tL):\|\N\|_2\leq 1}  \|\N\|_{\infty}$
are used to characterize the properties of the tangent spaces.
\begin{remark}
Our definitions of $\mu(\tS)$ and $\xi(\tL)$ are different from the respective definitions in \cite{CSPW_siam_2011}. In fact, the values of our $\mu$ and $\xi$ are less than or equal to the respective values in \cite{CSPW_siam_2011}. 
\end{remark}

The following proposition provides a sufficient condition to obtain the unique decomposition.

\begin{proposition}
\label{prop:gamma_suff_muxi}
Suppose that $\C=\tS+\tL$ is given. Suppose that
$ \mu(\tS)\xi(\tL)<\frac{1}{6}.$
Then, the unique optimum for \eqref{eq:convex_opti} is $(\hS_{\g},\hL_{\g})=(\tS,\tL)$ if
{\footnotesize \begin{align}
\label{eq:gamma_range_muxi}
    \g \in \left(\frac{\xi(\tL)}{1-4\mu(\tS)\xi(\tL)}, \frac{1-3\mu(\tS)\xi(\tL)}{\mu(\tS)} \right).
\end{align}
}
\end{proposition}
\hspace{-0.65cm}\begin{proof}
See supplementary material\black{, Appendix C} or \cite{doddi2019topology}. 
\end{proof}
\begin{remark}
The range of values in \eqref{eq:gamma_range_muxi} is a superset of the range specified in \cite{CSPW_siam_2011}. In the worst case, \eqref{eq:gamma_range_muxi} would return the same interval specified in \cite{CSPW_siam_2011}, due to difference in definitions of the tangent spaces, $\Omega(\tS)$ and  $T(\tL)$.
\end{remark}

 Next, we define $deg_{max}$ of a matrix $\M$ as
{\footnotesize
$deg_{max}(\M):=\max\left(\max_{1\leq i\leq n} \left( \sum_{j=1}^n \mathbbm{1}_{\{\M_{ij}\neq 0\}}\right)\right.,
\max_{1\leq j\leq n} \left.\left( \sum_{i=1}^n \mathbbm{1}_{\{\M_{ij}\neq 0\}}\right)\right),$
}where $\1_{\{x\neq 0\}}:=1$ if $x\neq 0$ and $\1_{\{x\neq 0\}}:=0$ if $x=0$ denotes the indicator function.
We define the \emph{maximum incoherence} of the row/column space of the real skew symmetric matrix $ \M$ as
$inc(\M):=\max_k \|\U \U^T e_k\|_2,$ where $\U\Sigma\V^T$ is the CSVD of $\M$. This definition is different from the one in \cite{CSPW_siam_2011} due to Lemma \ref{lem:proj_UV}.

The following lemma extends the sufficient condition in Proposition \ref{prop:gamma_suff_muxi} in terms of $deg_{max}$ and maximum incoherence.

\begin{lemma}{}
\label{lem:suff_gamma_deg_inc}
Let $\C= \tS + \tL$ with $deg_{max}(\tS)$ and $inc(\tL)$ as defined above. If
$deg_{max}(\tS)inc(\tL) <\frac{1}{12}$, then the unique optimum of the convex program \eqref{eq:convex_opti} is $(\widehat{\S}_{\g},\widehat{\L}_{\g})=(\tS,\tL)$ for a range of values of $\gamma$ given by:
{\footnotesize
	\begin{equation}
	\label{eq:gamma_range}
	\gamma \in \left( \frac{2 inc(\tL)}{1-8 deg_{max}(\tS) inc(\tL)}, \frac{1-6deg_{max}(\tS) inc(\tL)}{deg_{max}(\tS)}\right).
	\end{equation}
}\end{lemma}
\hspace{-0.65cm}\begin{proof}
The proof is similar to Corollary $3$ in \cite{CSPW_siam_2011}, and is skipped due to space constraint.
\end{proof}

Thus, by picking a proper $\gamma$ the convex optimization \eqref{eq:convex_opti} returns the unique decomposition $(\tS,\tL)$ without the need to determine $\O(\tS)$ and $T(\tL)$.

\begin{remark}
\label{rem:alg_matrix_decomp}	Lemma \ref{lem:suff_gamma_deg_inc} provides a conservative sufficient condition and hence covers only a subclass of uniquely decomposable matrices. In Section \ref{sec:simulation} we provide an example that does not satisfy the sufficient conditions, but still is uniquely decomposable using Algorithm \ref{alg:matrix_decomposition}, i.e., the network satisfies the transverse intersection \eqref{eq:transverse_intersection}, but not the sufficient condition in Lemma \ref{lem:suff_gamma_deg_inc}.	
\end{remark}
	
It can be observed from the structure of singular vectors in the projection matrix $\U\U^T$ that $n$ must be large for $inc(\widetilde{\L})$ to be small enough to satisfy the condition in Lemma \ref{lem:suff_gamma_deg_inc}. 
Moreover, using the results from \cite{robust_Matrix_decomposition}, it can be shown that the number of non-zero entries in $\widetilde{\S}$ must be at most $O(n)$ for the sufficient conditions to hold. 

The convex program \eqref{eq:convex_opti} is equivalent to the following formulation with the mapping $t=\frac{\g}{1+\g}$, where $t\in[0,1]$:
{\footnotesize
\begin{eqnarray}
\label{eq:convex_opti_t}
(\widehat{\S}_t,\widehat{\L}_t)&=& \arg \min_{\S,\L} t \|\S\|_1 + (1-t)\|\L\|_{*}\\
\nonumber
&& \text{subject to } \S+\L=\C,\\
&& \nonumber \hspace{1.53cm} \S^T=-\S, \ \L^T=-\L.
\end{eqnarray}	
}The following definitions are used to measure the closeness of the estimated matrices with the true matrices. 
{\footnotesize
\begin{eqnarray}
tol_{t}&:=&\frac{\|\widehat{\S}_{t}-\tS\|_F}{\|\tS\|_F}+\frac{\|\widehat{\L}_{t}-\tL\|_F}{\|\tL\|_F},\\
\label{eq:diff_t}
{diff}_t&:=&(\| \widehat{\S}_{t-\epsilon}-\widehat{\S}_{t} \|_F)+(\| \
\widehat{\L}_{t-\epsilon}-\widehat{\L}_{t} \|_F),
\end{eqnarray}
}where $\|.\|_F$ denotes the Frobenius norm and $\epsilon>0$ is a sufficiently small fixed constant.
Note that $tol_{t}$ requires the knowledge of the true matrices $\tS$ and $\tL$, whereas ${diff}_t$ does not require any such prior information. Moreover, the mapping between $t$ and $\g$ is one-to-one.

In practice, we may not have access to any extra information other than $\C$; thus determining $t$ required for the unique decomposition from Lemma \ref{lem:suff_gamma_deg_inc} and $tol_{t}$ becomes difficult. Here we provide guidance on which $t$ (and this $\gamma$) to be employed.
The following proposition provides a systematic approach to identify a proper penalty factor $t$ for the unique decomposition.
\begin{proposition}
	\label{prop:diff_t}
	Suppose we are given a matrix $\C$, which is obtained by summing $\tS$ and $\tL$, where $\tS$ is a sparse matrix and $\tL$ is a low-rank matrix. If $\tS$ and $\tL$ satisfies $\deg_{\max}(\tS) inc(\tL)<1/12$, then there exist at least three regions where $diff_t=0$. In particular, there exists an interval $[t_1,t_2]\subset[0,1]$ with $0<t_1<t_2<1$ such that $(\hS_t,\hL_t)=(\tS,\tL)$ for any $t \in [t_1,t_2]$
\end{proposition}
\hspace{-0.65cm}\begin{proof}
See Appendix \ref{appendix:diff_t}.
\end{proof}

\begin{corollary}\label{cor:refResult1}
By solving (\ref{eq:convex_opti_t}) and calculating $diff_t$ for every $t\in \{\eps,2\eps,\dots,1\}$ we obtain $\widehat{\S}_t$ and $\widehat{\L}_t$ and the zero regions specified in Proposition \ref{prop:diff_t}, specifically $[t_1,t_2], \ 0<t_1<t_2<1$, where $diff_t=0$. For a given $t\in [t_1,t_2]$, if $\deg_{\max}(\hat{\S}_t) inc(\hat{\L}_t)<1/12$, then the decomposition is exact, that is, $tol_{t}=0$ and $ (\widehat{\S}_t,\widehat{\L}_t)=(\tS,\tL)$. \end{corollary}

\begin{remark}
Proposition \ref{prop:diff_t} and Corollary \ref{cor:refResult1} are applicable in any general sparse plus low-rank matrix decomposition, and is not restricted to skew symmetric matrix decomposition.
\end{remark}

\begin{remark}
Conversely, if there are only two zero regions, $t$ close to zero and $t$ close to $1$, then we can assert that it may not be possible to obtain unique decomposition with this approach. Simulation results show that $t \in [.26,.4] $ is a good region to look for $t$.
\end{remark}

Based on Proposition \ref{prop:diff_t} and Corollary \ref{cor:refResult1}, we propose Algorithm \ref{alg:matrix_decomposition} to obtain the unique decomposition, which returns the estimated sparse matrix $\hS$ and estimated low-rank matrix $\hL$.
{\color{black} \begin{remark}
\label{rem:TS_complex}
The aforementioned results can be extended to the space of complex Hermitian matrices also; however, we do not discuss them here.
\end{remark}

In the following section, we discuss some preliminaries of linear dynamical systems that are useful in understanding the rest of the article. In Section \ref{sec:moral_graph} and Section \ref{sec:top_rec}, we discuss how the matrix decomposition is extremely useful to reconstruct the moral graph/topology of a given linear dynamical model.}

\section{Linear Dynamical Systems}
\label{sec:LD_system}
Consider a linear dynamical system with $n$ interacting agents, each equipped with time-series measurements $(\widetilde{x}_i(t))_{t\in \mathbb{Z}}$, $i\in\{1,\dots,n\}$, governed by the following {\color{black}linear dynamical model (LDM)}:
	\begin{align}\label{eqn:time_domain_model} 
	{\widetilde{\x}}(k) &= \sum_{l=-\infty}^{\infty} \widetilde{\H}(l){\widetilde{\x}}(k-l)+{\widetilde{\e}}(k), 
	\end{align}
	where $\widetilde{\x}(k) = [\widetilde{x}_1(k),\cdots,\widetilde{x}_n(k)]^T$,  ${\widetilde{\e}}(k) = [\widetilde{e}_1(k),\cdots,\widetilde{e}_n(k)]^T$, and for $i,j \in \{1,2,\cdots,n\},~i\neq j,$ $\widetilde{e}_i(k)$ is a zero mean wide sense stationary (WSS) process uncorrelated with $\widetilde{e}_j(k)$. Additionally, the processes $\{\widetilde{x}_i(k),\widetilde{e}_i(k)\}_{i=1}^n$ are jointly WSS. Let $\H(z) = \mathcal{Z}[\widetilde{\H}(k)]$. Then, $\widetilde{\H}(l) \in \mathbb{R}^{n \times n}$ denotes the weighted adjacency matrix with diagonal entries $\widetilde{\H}_{ii}(l)=0$, $1\leq i \leq n $, $l \in \mathbb{Z}$, {\color{black}such that $\H$ is well posed, i.e., every entry of $(\I-\H(z))^{-1}$ is analytic on the unit circle, $|z|=1,~z\in \mathbb{C}$. An LDM is said to be topologically detectable if $\Phi_\e(z)$ is positive definite for every $|z|=1$.
} The above model can be represented using the following Transfer Function Model (TFM),
	\begin{align}\label{eqn:TF_model}
	\x(z) &= \H(z)\x(z)+\e(z), ~ z \in \mathbb C,
	\end{align}
{\color{black} where $\x(z) = \mathcal{Z}[{\wx(k)}]$ and $\e(z) = \mathcal{Z}[\widetilde{\e}(k)]$. In general, there may exist nodes whose observations are not available and remain hidden. These nodes that are not accessible are called hidden/latent/unobservable nodes.} 
$\mV_o$ denotes the set of observable nodes with cardinality $n_o$ and $\mV_h$ is the set of latent nodes with cardinality $n_h$. 
\vspace{-10pt}
\subsection{Graphical Representation}
	
The Linear Dynamic Graph (LDG) associated with the LDM (\ref{eqn:TF_model}) is defined as the directed graph $\mathcal{G}(\mV,\mE)$, where $\mV=\{1,2,\cdots,n\}$ and $\mE=\{(i,j)|{\H}_{ji}\neq 0\}$. Thus, there exists a directed edge $(i,j)$ from node $i$ to node $j$ in the LDG if and only if ${\H}_{ji}\neq 0 $. For a directed graph $\mathcal{G}(\mV,\mE)$, parent set of node $j$ is $\mP(j):= \{i| (i,j)\in \mE \}$, child set of node $j$ is $\mC(j):=\{ i| (j,i)\in \mE\}$ and spouse set of node $j$ is $\mathcal{S}(j):=\{i|i \in \mP(\mC(j))\}$. {\color{black}Nodes $i$ and $j$ are strict spouses if $i \in \mathcal{S}(j)$, $i \notin \mC(j)\cup \mP(j)$. The Markov Blanket of node $i$, denoted $kin(i):=\mathcal{C}(i)\cup\mathcal{P}(i)\cup\mathcal{S}(i).$} The moral or the kin graph, {\color{black}$kin(\mathcal{G}):=\{(i,j)\mid i \in kin(j),~i,j \in \mV \}$, where $(i,j)$ denotes an unordered pair. The topology of $\mathcal{G}$ is defined as an undirected graph $top(\mathcal{G}):=\{(i,j) \mid  i \in \mP(j)\cup \mC(j),~i,j \in \mV \}.$} 

Similarly, an LDG obtained by restricting the vertex set to the observed nodes is defined by $\mathcal{G}_o(\mV_o,\mE_o)$, where $\mE_o:=\{(i,j)|i,j\in \mV_o \text{ and } {\H}_{ji}\neq 0\}$. The topology among the observable nodes, $\mathcal{T}(\mV_o,{\mE}_o):=top(\mathcal{G}_o(\mV_o,\mE_o))$. {\color{black} We define an undirected edge set $\overline{\mE}_o:=\{(i,j) \mid (i,j)\in \mE_o \text{ or } (j,i) \in \mE_o,~ i<j\}$.} 
Similarly, the moral graph {\color{black} among the observable nodes is the undirected graph {\small $kin(\mathcal{G}_o):=\{(i,j)\mid i \in kin(j),~i,j \in \mV_o \}$}}. We define a path between nodes $i$ and $j$ in an undirected graph as a set of nodes $\{i,x_0,\dots,x_k,j\}$ where $\{(i,x_0),(x_1,x_2),\dots ,(x_{k-1},x_k)\} \subseteq \mE$. We define a path between nodes $i$ and $j$ in an undirected graph as a set of nodes $\{i,x_0,\dots,x_k,j\}$ where $\{(i,x_0),(x_1,x_2),\dots ,(x_{k-1},x_k)\} \subseteq \mE$. A directed path in a directed graph is a path between nodes $i$ and $j$ with the constraint that all the edges are directed from $i$ towards $j$. $d_{hop}(i,j)$ is defined for undirected graphs as the number of links between nodes $i$ and $j$ on the shortest path connecting $i$ and $j$. It can be shown that $d_{hop}(i,j)$ is a metric for undirected graphs. For a node $i \in \mV$, the degree of the node is $deg(i):=|\{j\in \mV: (i,j) \in \mE\}|$. Note that this definition is for undirected edges.

{\color{black} In the next section, we study properties of the IPSDM of a given LDM that are useful in topology/moral graph reconstruction.}
\section{Exact Reconstruction of Moral Graph of Observed Nodes and Markov Blanket of Latent Nodes}
\label{sec:moral_graph}
\subsection{Moral Graph Reconstruction under Complete Observability}
\label{subsec:moral_graph_rec}
In this part of the section we present {\color{black} some important preliminaries} for reconstruction of moral graphs from power spectral density matrices and methods on how to reconstruct moral graph under full observability of the network.

{\color{black}For the graph $\mathcal{G}(\mV,\mE)$, the power spectral density matrix (PSDM), $\Phi_{\mathbf{x}}(z)\in \mathbb{C}^{n \times n}$ is given by
\vspace{-10pt}
{\small \begin{equation}
\label{eq:psd}
\Phi_{\mathbf{\x}}(z):= \sum_{k=-\infty}^{\infty} \mathbb{E} \{ \wx(k)\wx^T(0)\} z^{- k},~ z\in \mathbb{C}, |z|=1.
\end{equation}\vspace{-10pt}}

The PSDM and the IPSDM of the dynamical system governed by \eqref{eqn:TF_model} can be respectively written as (see \cite{materassi_tac12})	$\Phi_{\mathbf{x}}(z)=(\I-\H(z))^{-1}	\Phi_{\mathbf{e}}(z)(\I-\H^*(z))^{-1}$ and
	\begin{equation}
	\label{eq:PSD_inv}
	\Phi_{\mathbf{x}}^{-1}(z)=(\I-\H^{*}(z))	\Phi^{-1}_{\mathbf{e}}(z)(\I-\H(z)).
	\end{equation}}
	Note that $\Phi^{-1}_{\mathbf{e}}(z)$ is a diagonal matrix, since $\widetilde{e}_i(k)$ is uncorrelated with $\widetilde{e}_\ell(k)$ for $i \neq \ell$.

The following lemma {\color{black}(Theorem 27 in \cite{materassi_tac12})} provides a sufficient condition to estimate the moral graph of $\mG(\mV,\mE)$ from the IPSDM. 

{\color{black}
\begin{lemma}	\label{lem:Wiener_filter}
Consider a well posed and topologically detectable LDM $(\H,\e)$ with the associated LDG $\mathcal{G}(\mV,\mE)$, described by \eqref{eqn:TF_model}, having full node observability. Let IPSDM of $\x$ be given by \eqref{eq:PSD_inv}. Then, $[\Phi^{-1}_{\mathbf{x}}]_{i,j}\neq 0$, $i\neq j$ implies $i \in kin(j)$. Moreover, the converse holds almost always.
\end{lemma}
 \begin{remark}
 In \cite{zorzi_AR_hidden} and the related works, $[\Phi^{-1}_{\mathbf{x}}]_{i,j}\neq 0$ is considered to be equivalent to nodes $i$ and $j$ being conditionally dependent given the rest of the observations. That is, retrieving conditional dependence is equivalent to reconstructing moral graph. 
\end{remark} 
\begin{remark}
{The results of \cite{zorzi_AR_hidden} can be extended to AR models with WSS noise by considering conditional correlation instead of conditional dependence. In this case, retrieving conditional correlation is equivalent to moral graph reconstruction.}
\end{remark}
}

\vspace{-5pt}
\subsection{Structure of the IPSDM with Latent Nodes}
	\label{subsec:top_S+Ldecomp}
In the previous subsection we studied the properties of IPSDM under the assumption that {\color{black}all the nodes are observable}. However, topology identification becomes complicated in the presence of latent nodes, \black{often leading to lack of identifiability} (see \ref{subsec:identifying_HN}). Here, we discuss some of the special properties of the IPSDM in the presence of latent nodes that are exploited in this article.
	
By separating observable nodes and latent nodes, we represent {\footnotesize$\x(z) =\begin{bmatrix}\mathbf{x}_o(z)\\\mathbf{x}_h(z)\end{bmatrix}$} and {\footnotesize$\e(z) =\begin{bmatrix}\mathbf{e}_o(z)\\\mathbf{e}_h(z)\end{bmatrix}$}, where {$\x_o(z) = \mathcal{Z}[{\wx_o(k)}]$, $\x_h(z) = \mathcal{Z}[{\wx_h(k)}]$, ${\e}_o(z) = \mathcal{Z}[{\widetilde{\e}_o(k)}]$, and $\e_h(z) = \mathcal{Z}[\widetilde{\e}_h(k)]$}. The TFM
	in (\ref{eqn:TF_model}) can be expressed as follows:
{\footnotesize
	\begin{align}
	\label{eq:TF_block_model}
\begin{bmatrix}\mathbf{x}_o(z)\\\mathbf{x}_h(z)\end{bmatrix}= \begin{bmatrix}\H_{oo}(z)&\H_{oh}(z)\\\H_{ho}(z)&\H_{hh}(z) \end{bmatrix} \begin{bmatrix}\mathbf{x}_o(z)\\\mathbf{x}_h(z)\end{bmatrix}+\begin{bmatrix}\mathbf{\e_o}(z)\\\mathbf{\e_h}(z)\end{bmatrix}.
	\end{align}	
}	
Letting 
{\footnotesize $	\Phi_{\mathbf{x}}(z)=\left[\begin{matrix}
	{\Phi}_{oo}(z) & {\Phi}_{oh}(z)\\ 
	{\Phi}_{ho}(z) & {\Phi}_{hh}(z)
	\end{matrix}\right]$}
	and {\footnotesize
$\Phi_{\mathbf{x}}^{-1}(z)=\left[\begin{matrix}
{\K}_{oo}(z) & {\K}_{oh}(z)\\ 
{\K}_{ho}(z) & {\K}_{hh}(z)
\end{matrix}\right]$} we have that (by ignoring the index $z$) $\Phi^{-1}_{oo}=\K_{oo}-\K_{oh}\K^{-1}_{hh}\K_{ho}$, which follows by applying block matrix inversion formula and using Schur’s complement representation \cite{matrix_analysis}. Furthermore, using (\ref{eq:PSD_inv}), the IPSDM corresponding to the observed variables can be written as:
	\begin{align}	\label{eq:S_L_decomp}
	\Phi_{oo}^{-1}=&\S+\L,\text{ where }\\
	\label{eq:S_def}
	\S&=(\I_o-\H_{oo}^{*})\Phi_{e_o}^{-1}(\I_o-\H_{oo}),\\
	\label{eq:L_def}
	\L &=\H_{ho}^{*}\Phi_{e_h}^{-1}\H_{ho}-\Psi^{*}\Lambda^{-1}\Psi,\\ \nonumber
	\Psi&=\H_{oh}^{*}\Phi_{e_o}^{-1}(\I-\H_{oo})+(\I-\H^*_{hh})\Phi_{e_h}^{-1}\H_{ho}, \text{ and}\\ \nonumber
	\Lambda&=\H_{oh}^{*}\Phi_{e_o}^{-1}\H_{oh}+(\I-\H^*_{hh})\Phi_{e_h}^{-1}(\I-\H_{hh}).\nonumber
	\end{align}
{\color{black} The following proposition shows that $support(\S)$ can retrieve the moral graph among the observable nodes, $kin(\mG_o)$.
\begin{proposition}
\label{prop:S_top_obs_rec}
Consider a well-posed and topologically detectable LDM, $(\H,\e)$, described by \eqref{eqn:TF_model}, with the associated graph $\mG(\mV,\mE)$. Let $\S$ be given by \eqref{eq:S_def}. Then, the set $\widehat{\mE}_o:=\{(i,j)| \S_{ij} \neq 0, i<j \}$ reconstructs $kin(\mG_o)$ almost always.
\end{proposition}
 \begin{proof}
Notice that \eqref{eq:PSD_inv} and \eqref{eq:S_def} are exactly the same, except $\H,~\Phi_{\e}$, and $\I$ in \eqref{eq:PSD_inv} are replaced with $\H_{oo},~\Phi_{\e_o}$, and $\I_o$ respectively in \eqref{eq:S_def}. Thus, by applying Lemma \ref{lem:Wiener_filter} on $\S$, $\widehat{\mE}_o$ retrieves  $kin(\mG_o)$ almost always.
\end{proof}

 The following theorem shows that $\S$ is sparse if the moral graph, $kin(\mG_o)$ is sparse, while $\L$ is a low-rank matrix if $n_h<<n_o$. This particular structure aids in decomposing $\Phi_{oo}^{-1}$ into $\S$ and $\L$ (or more precisely $\Im\{\Phi_{oo}^{-1}\}$ into $\Im\{\S\}$ and $\Im\{\L\}$ using the results from Section \ref{sec:matrix_decomp}). Notice that the index $z$ is omitted from the notations. The results hold for every $ |z|=1$ uniformly.
	
\begin{theorem}
\label{theorem:S_L_structure}
Consider the LDG $\mathcal{G}(\mV,\mE)$ described by \eqref{eqn:TF_model}. Let $\mG_o(\mV_o,\mE_o)$ be the LDG, $\mathcal{G}(\mV,\mE)$, restricted to the observed nodes. The following holds:  

\vspace{-10pt}
{\small \begin{eqnarray}
\label{eq:thm1_sparse}
|support(\S)|\leq2|\mE_o|+ 2|\mE^{ss}_o|+n,\ 
rank(\L) \leq 2 n_h,
\end{eqnarray}}
\normalsize
\noindent 
where $\mE^{ss}_o$ denotes the set of undirected edges between the observable strict spouses with a common observable child.
\end{theorem}
\hspace{-0.65cm}\begin{proof}
See Appendix \ref{appendix:sparse_LR}.
\end{proof}


\begin{remark}
\label{rem:thm1}
Here, we are interested in the scenarios where $2|\mE_o| +2|\mE_o^{ss}|=O(n)$; note that the maximum number of interconnections in a graph of $n$ nodes is $n^2$. In this sense, $\S$ is considered {\it sparse}. If we can uniquely decompose $\Phi_{oo}^{-1}(z)$ into the sparse matrix $\S$ and the low-rank matrix $\L$, then one can obtain $kin(\mG_o)$ from $\S$ (see Proposition \ref{prop:S_top_obs_rec}) and the Markov Blanket of the hidden nodes from $\L$ (see Section \ref{subsec:markov_blanket}). Further, in Section \ref{subsec:top_rec}, we reconstruct the exact topology of the entire network, including that of hidden nodes, under some assumptions. However, there are certain identifiability issues related to hidden nodes, inherent in network topology, which make the detection of hidden nodes from $\L$ difficult, even impossible in some cases. We address them next.
 \end{remark}
}

\subsection{Identifiability of Latent Nodes}
\label{subsec:identifying_HN}
	
{\color{black} Here, we discuss identifiability of the hidden nodes inherent to the graph structure and not limited to any specific reconstruction method.} We illustrate the non-identifiability via examples.

\begin{figure}
\centering
\includegraphics[width=0.9\columnwidth]{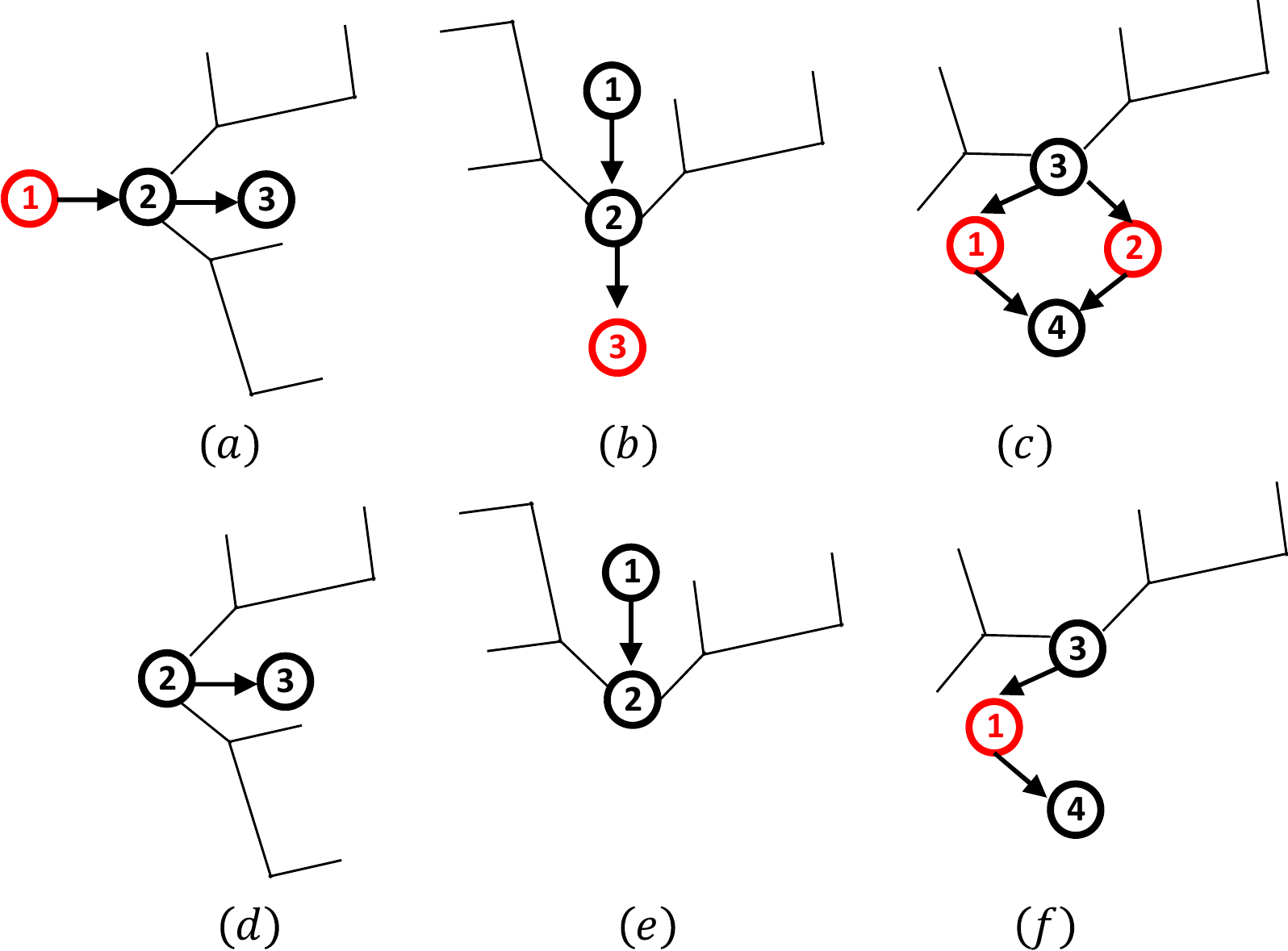}
\caption{Non-identifiability of hidden nodes (red colored nodes): (a) hidden node is terminal and strict parent (b) hidden node is terminal and strict child (c) Markov Blankets of hidden nodes $1$ and $2$ overlapped.}
\label{fig:conn_id}
\end{figure}

Fig. \ref{fig:conn_id}a and Fig. \ref{fig:conn_id}d represent {\color{black} LDG}s with LDMs given by 
$\widetilde{x}_1=\widetilde{e}_1,\ 
\widetilde{x}_2 = \widetilde{x}_1+\sum_{i=4}^{n}h_{2i}*\widetilde{x}_i(k) +\widetilde{e}_2,\ 
\widetilde{x}_3 = h_{32}*\widetilde{x}_2(k) +\widetilde{e}_3$, where node 1 is latent, and $    \widetilde{x}_2=\sum_{i=4}^{n}h_{2i}*\widetilde{x}_i(k)+\hat{e}_2 ,\ 
\widetilde{x}_3 = h_{32}*\widetilde{x}_2(k) +\widetilde{e}_3$ respectively. Let $ \hat{e}_2=\widetilde{e}_2+\widetilde{e}_1$. Then the observed node time-series \black{obtained from the two LDMs} are identical and the two models are indistinguishable from $\Phi_{oo}^{-1}$.

Similarly, consider the {\color{black} LDG}s shown in Fig. \ref{fig:conn_id}b and Fig. \ref{fig:conn_id}e with \black{LDM}s given by 
$\widetilde{x}_1=\widetilde{e}_1,\ 
\widetilde{x}_2 = \widetilde{x}_1+\sum_{i=4}^{n}h_{2i}*\widetilde{x}_i(k) +\widetilde{e}_2, \ \widetilde{x}_3 = h_{32}*\widetilde{x}_2(k) +\widetilde{e}_3$, $   \widetilde{x}_1=\widetilde{e}_1,\ 
\widetilde{x}_2 = \widetilde{x}_1+\sum_{i=4}^{n}h_{2i}*\widetilde{x}_i(k) +\widetilde{e}_2$ respectively, where node 3 is latent. Again, the time-series among the observable nodes {\color{black}obtained from both the LDG}s are the same and hence the two models are indistinguishable.
    
Based on the aforementioned discussion, we make the following assumptions for identifiability of a hidden node.
\begin{assump} \label{ass:atleastChild}
Any hidden node $k_h$ in $\mG(\mV,\mE)$ has at least one observed-child $c\in \mV_o$. Further, $k_h$ is a parent or
child of another observable node $j\in \mV_o\setminus\{c\}$.  
\end{assump}

We next illustrate that non-identifiability issues arise when the Markov Blankets of hidden nodes overlap.
Consider the {\color{black} LDG}s shown in Fig. \ref{fig:conn_id}c and Fig. \ref{fig:conn_id}f with \black{LDM}s given by $\widetilde{x}_3=\sum_{i=5}^{n}h_{3i}*\widetilde{x}_i(k) +\widetilde{e}_3,\     \widetilde{x}_1=\widetilde{x}_3+\widetilde{e}_1, \ 
\widetilde{x}_2=\widetilde{x}_3+\widetilde{e}_2, \ \widetilde{x}_4=\widetilde{x}_1+\widetilde{x}_2+\widetilde{e}_4,$
and $\widetilde{x}_3=\sum_{i=5}^{n}h_{3i}*\widetilde{x}_i(k) +\widetilde{e}_3,\
\widetilde{x}_1=\widetilde{x}_3+\hat{e}_1, \ \widetilde{x}_4=\widetilde{x}_1+\widetilde{e}_4,$ where $\hat{e}_1=\widetilde{e}_1+\widetilde{e}_2$ , respectively, where nodes 1 and 2 are latent. Both \black{LDM}s result in the same observed time-series, which leads to non-identifiability of hidden node $2$. Hence, we make the following assumption about the spatial distribution of hidden nodes in the {\color{black} LDG} $\mG(\mV,\mE).$
\begin{assump}
\label{ass:latent5hop}
For every distinct $k_h,k_h' \in \mV_h$, $d_{hop}(k_h,k_h')>4$.
\end{assump}

Assumption $\ref{ass:latent5hop}$ is sufficient to ensure that Markov Blankets of any two distinct hidden nodes $k_h$ and $k_h'$ do not overlap. When the intersection of the Markov Blankets of $k_h$ and $k_{h'}$ contain more than one observable node, then one of the hidden nodes $k_h$ or $k_{h'}$ is non-identifiable (see the illustration associated with Fig. \ref{fig:conn_id}c). If the intersection has at most one node, then both the hidden nodes {\color{black} might be} identifiable. We make a slightly more conservative assumption that the Markov Blankets of the hidden nodes are non-overlapping. Moreover, the implication of Assumption \ref{ass:latent5hop} is that the Markov Blanket of a hidden node $h\in \mV_h$ is $\{\mathcal{P}(h)\cup \mathcal{C}(h)\cup \mathcal{S}(h)\} \subseteq \mV_o.$ 

\begin{remark}
Note that Assumption \ref{ass:latent5hop} might seem stronger than Assumption 2 in \cite{Talukdar_ACC18}. However, \cite{Talukdar_ACC18} restricted attention to radial topologies associated with bi-directed {\color{black} LDG}s and assumed that the hidden nodes are at least three hops away from the leaf nodes. On the contrary, algorithms in {\color{black} Section \ref{sec:top_rec} can reconstruct more general linear dynamical networks, including loopy networks.}
\end{remark}

\subsection{Markov Blanket Reconstruction of the Latent nodes}
\label{subsec:markov_blanket}

Here, we provide the following definition and theorems which helps in learning the Markov Blanket of hidden nodes. 
\black{\begin{definition}\label{def1}
For $i,j\in \mV_o$ and hidden node $h\in \mV_h,$ define {\small$DE_h(i,j):= \{g_1,g_2,g_3,g_4,g_5,g_6,g_7,g_8,g_9\},$ $DE_h^{ss}(i,j):=\{g_{10},g_{11}\}$}, and {\small$DM_h(i,j):=DE_h(i,j) \cup DE_h^{ss}(i,j)$}, where {\small$i\rightarrow h$} is denoted as $(i,h)$ with {\footnotesize$g_1  = \{(h,i),(h,j)\}$, $g_2 = \{(h,i),(j,h)\}$, $g_3 = \{(h,i),(h,k),(j,k)\},\ \forall k\in \mV_o\setminus\{i,j\}$, $g_4 = \{(i,h),(h,j)\}$, $g_5 = \{(i,h),(h,k),(j,k)\},\ \forall k\in \mV_o\setminus\{i,j\}$, $g_6 = \{(i,k),(h,k),(j,h)\},\ \forall k\in \mV_o\setminus\{i,j\}$, $g_7 = \{(i,k),(h,k),(h,j)\},\ \forall k\in \mV_o\setminus\{i,j\}$, $g_8 = \{ (h,i),(j,i) \},\  g_9 = \{ (h,j),(i,j) \},$ $g_{10} = \{(i,h),(j,h) \},$ and $g_{11}= \{(i,k_1),(h,k_1),(h,k_2),(k_2,j) \},\ \forall k_1,k_2 \in \mV_o \setminus \{ i,j\}.$ }
\end{definition}
$DM_h(i,j)$} enumerates all the possible paths between two observable nodes 
$i,j$ present in the Markov Blanket of hidden node $h.$ We now present a result which infers the Markov Blanket of a hidden node $h$ in $\mG(\mV,\mE)$ from $support(\L).$ 
\begin{theorem}\label{thm1} {\color{black}Suppose the LDM in \eqref{eq:TF_block_model} satisfies assumptions \ref{ass:atleastChild},~\ref{ass:latent5hop}. Let $\L$ be given by \eqref{eq:L_def} and let $i,j\in \mV_o,~ i\neq j.$} Then, the following statements hold:
\begin{enumerate}[label=(\alph*)]
\item If ${{\color{black}\L_{ij}}}\neq 0,$ then there exists $g \in DM_h(i,j)$ such that $g \in \mG(\mV,\mE),$ for some $h\in \mV_h$. Further, {\color{black}$\dh({i,h})\leq 2$ and $\dh({j,h})\leq  2.$}
\item Given ${{\color{black}\L_{ij}}}\neq 0,$ suppose there exist $g_1\in DM_{h_1}(i,j)$ and $g_2\in DM_{h_2}(i,j)$ connecting $i$ and $j$ such that $g_1,g_2 \in \mG(\mV,\mE)$, for some $h_1,h_2\in \mV_h.$ Then $h_1=h_2.$
\end{enumerate}
\end{theorem}

\hspace{-0.65cm}\begin{proof}
See Appendix \ref{appendix:thm1}.
\end{proof}

\begin{remark}\label{rem:pathological}
We note that, for a set of system parameters, noise statistics can be construed such that a $g\in DM_h(i,j)$ is present in the {\color{black} LDG} $\mG(\mV,\mE)$, with ${{\color{black}\L_{ij}}}= 0.$ We remark that such cases are pathological; we will assume that the converse of Theorem \ref{thm1}(a) holds almost everywhere.
\end{remark}

Thus, based on the locations of non-zero entries in $\L,$ we construct $V_H = \{i | i\in \mV_o,\exists k\in \mV_o\setminus\{i\}\text{ s.t. }{{\color{black}\L_{ik}}} \neq 0 \}$ and 
$E_H = \{(i,j) | i,j\in V_H,{{\color{black}\L_{ij}}} \neq 0 \}.$ The following result shows that $(V_H,E_H)$ is a disjoint collection of connected undirected subgraphs (see Fig. \ref{fig:top_hidden} for example). Moreover, the number of connected undirected subgraphs in $E_H$ is equal to number of hidden nodes $n_h$ in the {\color{black} LDG} $\mG(\mV,\mE).$

\begin{theorem} \label{thm2}
{\color{black}Suppose the LDM in \eqref{eq:TF_block_model} satisfies assumptions \ref{ass:atleastChild},~\ref{ass:latent5hop}. Let $\L$ be given by \eqref{eq:L_def}. From} ${\L},$ construct the following:
$V_H = \{i | i\in \mV_o,\exists k\in \mV_o\setminus\{i\}\text{ s.t. }{{\color{black}\L_{ik}}} \neq 0 \},$
$E_H = \{(i,j) | i,j\in V_H,{\L_{ij}} \neq 0 \}.$ For every hidden node $l\in \mV_h,$ let $M_l =\mathcal{P}(l)\cup \mathcal{C}(l)\cup \mathcal{S}(l),Q_l = \{(i,j)| i,j \in M_l,{\L_{ij}} \neq 0\}.$ Then 
\begin{enumerate}[label=(\alph*)]
\item $M_{l_1}\bigcap M_{l_2}=\emptyset,$ for all $l_1,l_2 \in \mV_h, l_1\neq  l_2.$
\item $Q_{l_1}\bigcap Q_{l_2}=\emptyset,$ for all $l_1,l_2 \in \mV_h, l_1 \neq  l_2.$
\item $V_H = \bigcup\limits_{l=1}^{n_h} M_l.$
\item $E_H = \bigcup\limits_{l=1}^{n_h} Q_l.$
\end{enumerate}
\end{theorem}

\hspace{-0.65cm}\begin{proof}
See Appendix \ref{appendix:thm2}.
\end{proof}

\begin{remark}\label{rem:island}
The above theorem estimates the number of hidden nodes $n_h$ as the number of connected undirected subgraphs in $E_H.$ Each connected component $(M_l,Q_l)$ is due to a hidden node $l \in \mV_h,$ that is, $M_l$ {\color{black} is the Markov Blanket of $l$ in $\mG(\mV,\mE)$, while $Q_l$ is an undirected edge set that contain edges among any two distinct nodes in $M_l$.}
\end{remark}

{\color{black} 
\subsection{Moral Graph Reconstruction of Observable Nodes and Localization of Hidden Nodes from $\Phi_{oo}^{-1}$}
 \label{subsec:matrix_decomp_complex}

As shown in Theorem \ref{theorem:S_L_structure}, $\S$ is sparse and $\L$ is low-rank. Then, one can retrieve $\S$ and $\L$ from $\Phi_{oo}^{-1}$ by employing the following optimization.
{\footnotesize
\begin{eqnarray}
\label{eq:convex_opti_complex}
(\widehat{\S}_{\g},\widehat{\L}_{\g})&=& \arg \min_{\S,\L \in \mathbb C^{n \times n}} \g \|\S\|_1 + \|\L\|_{*}\\
\nonumber
&& \text{subject to } \S+\L=\Phi_{oo}^{-1},\\
\nonumber
&& \hspace{1.53cm} \S-\S^*=\mathbf{0}, \ \L-\L^*=\mathbf{0}.
\end{eqnarray}}
Section \ref{sec:matrix_decomp} has provided certain sufficient conditions for the unique decomposition, in the space of skew symmetric matrices. As mentioned in Remark \ref{rem:TS_complex}, one can extend the results to the space of complex Hermitian matrices also, which can be applied to solve \eqref{eq:convex_opti_complex}. Then, $support(\hS_{\gamma})$ recovers the exact moral graph among the observable nodes, $kin(\mG_o)$, and $rank(\hL_{\gamma})$ provides a lower bound on the number of hidden nodes. Additionally, as shown in Theorem \ref{thm2}, $support(\hL_{\gamma})$ retrieves Markov Blankets of all the hidden nodes.


\subsection{Reconstruction based on $\Im\{ \Phi_{oo}^{-1}(z)\}$}
In Theorem \ref{theorem:S_L_structure}, it was shown that $\S$ is sparse and $\L$ is low-rank, which implies that the same applies respectively to $\Im\{\S\}$ and $\Im\{\L\}$. Based on Theorem \ref{theorem:S_L_structure}, one can show that ${\small|support(\Im\{\S\})|\leq 2|\mE_o|+ 2|\mE^{ss}_o}|$ and
${\small rank(\Im\{\L\}) \leq 4 n_h}$. Therefore, $\Im\{\S\}$ is sparse and $\Im\{\L\}$ is low-rank.  As $\Phi_{oo}^{-1}$ is Hermitian, $\Im\{\Phi_{oo}^{-1}\}$ is skew symmetric, and hence the results from Section II are applicable here. Then, by applying the convex optimization \eqref{eq:convex_opti_t} with $\C=\Im\{\Phi_{oo}^{-1}(z)\}$, one can retrieve the ground truth $\Im\{\S(z)\}$ and $\Im\{\L(z)\}$, for all $|z|=1$, and appropriately selected $\g$ as shown in Section \ref{sec:matrix_decomp}. This procedure is provided in Algorithm 1. 
The caveat of decomposing $\Im\{ \Phi_{oo}^{-1}\}$ is that the following assumption is required for consistent moral graph/topology reconstruction from $\Im\{\S\}$ and $\Im\{ \L\}$.
 \begin{assump}
\label{ass:H_real_const}
For any $1 \leq i,k \leq n$, $i\neq k$, if $\H_{ik}(z)\neq0,$ then $\Im\{\H_{ik}(z)\}\neq 0$, for all $z, ~|z|=1$.
\end{assump}

\begin{remark}
Assumption 3 is necessary to reconstruct $kin(\mG_o)$ and the Markov Blankets of the hidden nodes from the decomposition of $\Im\{\Phi_{oo}^{-1}\}$, instead of decomposing $\Phi_{oo}^{-1}$ directly. From (\ref{eq:S_def}), (\ref{eq:L_def}), it follows that $\Im\{\S\}$ and $\Im\{\L\}$ depend on elements of $\Im\{\H\}$. Hence, when Assumption \ref{ass:H_real_const} holds, the Lemma \ref{lem:Wiener_filter}, Proposition \ref{prop:S_top_obs_rec}, Theorem \ref{theorem:S_L_structure}, \ref{thm1}, and \ref{thm2} hold by replacing $\Phi_x^{-1},~\Phi_{oo}^{-1}, ~\S$, and $\L$ with $\Im\{\Phi_x^{-1}\}, ~\Im\{\Phi_{oo}^{-1}\} , ~\Im\{\S\}$, and $\Im\{\L\}$ respectively. In other words, $kin(\mG_o)$ and the Markov Blanket of hidden nodes can be obtained by decomposing $\Im\{\Phi_{oo}^{-1}\}$ instead of $\Phi_{oo}^{-1}$. We focus on the decomposition of $\Im\{\Phi_{oo}^{-1}\}$ in this article.\end{remark} 


 
 


\begin{remark} \label{rem:spurious}
$kin(\mathcal{G}_o)$, obtained from $\Im\{\S\}$ using Proposition \ref{prop:S_top_obs_rec}, would contain $top(\mathcal{G}_o)$ as well as additional edges due to strict spouses in $\mathcal{G}_o.$ Likewise, $(M_l, Q_l)$ obtained from $support(\Im\{\L\})$ using Theorem \ref{thm2} may contain edges apart from $P(l) \cup C(l)$ of a hidden node $l$. Such spurious edges maybe many; examples include bi-directed \black{LDGs}. There is a need to eliminate them so that exact recovery of $top(\mathcal{G})$ is possible. 
\end{remark}

In the next section, we develop techniques to eliminate the spurious edges and reconstruct the exact topology, for a wide class of networks.

\vspace{-4pt}
}

\section{Exact Topology Reconstruction}	
\label{sec:top_rec}


{\color{black} 
In this section, we develop methods for exact recovery of $top(\mG)$, under Assumption \ref{ass:diff_eqn}, which is applicable to wide class of applications.

\vspace{-5pt}
\subsection{Elimination of Strict Spouse Edges}
Here, we show that, in certain LDMs, the strict spouse edges satisfy properties that can be exploited toward exact topology reconstruction. Notice that for \emph{some} of the results (viz Theorem \ref{theorem:S_top_obs_rec} and Theorem \ref{theorem:two_Par_Sp}), we restrict our interest to the models that satisfy the following assumption. For the networks that satisfy Assumption \ref{ass:diff_eqn}, Theorem \ref{theorem:S_top_obs_rec} will show that $support(\Im \{ \S \})$ would reconstruct the exact topology, $top(\mG_o)$ (not $kin(\mG_o)$). 
\begin{assump}
\label{ass:diff_eqn}
For the LDM in \eqref{eqn:TF_model}, and $i,k,l \in \mV$, if $\H_{ki}(z) \neq 0$ and $\H_{kl}(z) \neq 0$, then $\phase{\H_{ki}(z)}=\phase{ \H_{kl}(z)}$. 
\end{assump}
\begin{remark}
Assumption \ref{ass:diff_eqn} is satisfied by a large class of engineering systems. For instance, Section $2$ of \cite{TALUKDAR_physics} provides engineering systems that satisfy Assumption \ref{ass:diff_eqn}. Also, other examples include linearized chemical reaction ODEs \cite{2021Mathematics}. For example, see (10.2.2) in \cite{2021Mathematics}. 
\end{remark}

The following lemma (Theorem 3 in \cite{TALUKDAR_physics}) is useful in proving the subsequent results and helps in exploiting additional structure enjoyed by the IPSDM in the networks satisfying Assumption \ref{ass:diff_eqn}.

\begin{lemma}
\label{lem:spurious_edges}
Consider a well-posed and topologically detectable LDM $(\H,\e)$ with the associated graph $\mathcal{G}(\mV,\mE)$, described by \eqref{eqn:TF_model}, having full node observability and satisfying Assumption \ref{ass:diff_eqn}. If $i$ and $j$ are strict spouses, then 
$[\Im\{\Phi_{\x}^{-1}\}]_{ij}=0.$
\end{lemma}
}

\black{Lemma \ref{lem:spurious_edges} eliminates the spurious edges formed due to strict spouse connections by observing the entries of $\Im\{\Phi_x^{-1} \}$. The proof follows from the expansion of (\ref{eq:PSD_inv}). Here, $[\Im\{ \Phi_x^{-1}\}]_{ij} =0$ if $\H_{ij}=0,$ $\H_{ji}=0,$ and $\phase{\H_{ki}} = \phase{\H_{kj}}$ for all $k \in \mC(i) \cap \mC(j).$} This can be employed to separate the true parent-child connection from the strict spouse edges in the moral graph \black{obtained from Lemma \ref{lem:Wiener_filter}}.
{\color{black} Combining Lemma \ref{lem:Wiener_filter} and Lemma \ref{lem:spurious_edges} with Assumption \ref{ass:H_real_const}, one can conclude that, the undirected graph constructed from $\Im\{\Phi_x^{-1}\}$ is equal to $top(\mG)$, if Assumption \ref{ass:diff_eqn} holds; else it would be $kin(\mG)$.
We use this fact in the following theorem to reconstruct $top(\mG_o)$.
\begin{theorem}
\label{theorem:S_top_obs_rec}
Consider a well-posed and topologically detectable LDM $(\H,\e)$ with the associated graph $\mG(\mV,\mE)$, described by \eqref{eqn:TF_model} and satisfying Assumption \ref{ass:diff_eqn}. Let $\S$ be given by \eqref{eq:S_def} and let $\widehat{\mE}_o:=\{(i,j):\Im\{\S_{ij}\}\neq 0,~i<j\}$. Then, $\widehat{\mE}_o\subseteq \overline{\mE}_o$. Additionally, if the LDM satisfies Assumption \ref{ass:H_real_const}, then $\hat{\mathcal{E}}_o=\overline{\mathcal{E}}_o$ almost always. 
\end{theorem}}
\hspace{-0.65cm}\begin{proof}
See Appendix \ref{app:theorem:S_top_obs_rec}.
\end{proof}	
In the following, we discuss the exact topology reconstruction from $\Im\{\Phi_{oo}^{-1}\}$ based on Theorems \ref{thm2} and \ref{theorem:S_top_obs_rec}.

\begin{algorithm}
\caption{Matrix decomposition}
\textbf{Input:}$\Phi_{oo}^{-1}(z)$: IPSDM among $\mV_o$, $\varepsilon$,  \black{$z=e^{j\omega},~\omega \in (-\pi,\pi]$}\\
		\textbf{Output:} Matrices $\Im(\S(z))$ and $\Im(\L(z))$ 
		\label{alg:matrix_decomposition}
		\begin{algorithmic}[1]
		\State Set $\C=\Im\{\Phi_{oo}^{-1}(z)\}$ \State Initialize $(\widehat{\S}_{0},\widehat{\L}_{0})=(\C,\mathbf{0})$
			
			\ForAll{$t \in \{\eps,2\eps,\dots,1\}$}
			\State Solve the convex optimization \eqref{eq:convex_opti_t} and calculate ${diff}_t$ in \eqref{eq:diff_t}
			\EndFor
			\State Identify the three regions where $diff_t$ is zero and denote the middle region as $[t_1,t_2]$. 
			Pick a $t_0 \in [t_1,t_2]$ and the corresponding pair $(\hat{\S}_{t_0},\hat{\L}_{t_0})$.
\If{$deg_{max}(\hat{\S}_{t_0})inc(\hat{\L}_{t_0})<\frac{1}{12}$}

\State {\color{black}$(\widehat{\S}(z),\widehat{\L}(z))=(\hS_{t_0},\hL_{t_0})$}		
\State {\color{black}Return $(\widehat{\S}(z),\widehat{\L}(z))$ }
\EndIf 
\end{algorithmic}
\end{algorithm}

\subsection{Reconstruction of $\mT(\mathcal{\mV},\mE)$:}
\label{subsec:top_rec}
The topology, $\mT(\mathcal{\mV},\mE)$, of the {\color{black} LDG} $\mG(\mV,\mE)$ will be reconstructed in three steps: (a) recover the topology restricted to observed nodes given by $\mT(\mV_o,\mE_o),$ (b) determine the number of hidden nodes $n_h$ in the {\color{black} LDG} $\mG(\mV,\mE)$ and (c) reconstruct the topology associated with each hidden node. We denote the reconstructed topology as $\mT(\mV_R,\mE_R).$ The reconstruction is said to be exact when $\mT(\mV,\mE)= \mT(\mV_R,\mE_R).$

{\color{black} 
The first step in topology reconstruction is to obtain $\hS=\Im\{\S(z)\}$ and $\hL=\Im\{\L(z)\}$ from Algorithm \ref{alg:matrix_decomposition}. Then, from $\hS$, we reconstruct the topology among observed nodes $\mT(\mV_o,\mE_o)$ as shown in Theorem \ref{theorem:S_top_obs_rec}. This is performed in Algorithm \ref{alg:top_rec}, which reconstructs $\mathcal{T}(\mV_o,\mE_R)$, where 
$\mV_o$ is the set of observable nodes and $\mE_R$ is the undirected edge set reconstructed from $support(\hS)$.
Apart from pathological cases, $\mathcal{T}(\mV_o,\mE_R)$ is identical to $\mT(\mV_o,\mE_o)$ under Assumption \ref{ass:diff_eqn}.
\begin{remark}
Algorithms \ref{alg:matrix_decomposition} and \ref{alg:top_rec} are applicable for any LDM with Assumption \ref{ass:H_real_const} and need not satisfy Assumption \ref{ass:diff_eqn}. However, in that case, $\mathcal{T}(\mV_o,\mE_R)$ would return $kin(\mG_o)$.
\end{remark}
}
{\color{black} 
\begin{remark}
As mentioned in Section \ref{subsec:matrix_decomp_complex}, one can decompose the complex $\Phi_{oo}^{-1}$ to obtain $\S$ and $\L$, instead of their respective imaginary parts, thus avoiding the need for Assumption \ref{ass:H_real_const}. $\S$ and $\L$ can then retrieve moral graphs among the observable nodes and Markov Blanket of hidden nodes.
\end{remark}}

{\tiny
\begin{algorithm}
\caption{Topology reconstruction of observable nodes}
\textbf{Input:} $\hS(z)$ from Algorithm \ref{alg:matrix_decomposition},  threshold $\tau$\\
\textbf{Output:} Reconstructed topology among observable nodes $\mT(\mV_o,\mE_R)$
\label{alg:top_rec}
\begin{algorithmic}[1]
	\State Edge sets ${\mathcal{E}}_R \gets \{\}$
	\ForAll{$(i,j) \in \{1,2,...,n\} \bigtimes \{1,2,...,n\}$}
	\If{$|\hS_{ij}(z)| >\tau$}
	\State ${\mathcal{E}}_R \gets {\mathcal{E}}_R \cup \{(i,j)\}$
	\EndIf
	\EndFor
	\State $\mV_o \gets \{1,2,\cdots, n\}$
	\State Return $\mT(\mV_o,\mE_R)$
\end{algorithmic}
\end{algorithm}
}

We now proceed with estimating number of hidden nodes and reconstructing the topology associated with hidden nodes using $\Im\{\L\}$.

We emphasize that the undirected edges in $Q_l$ are not necessarily present in the true topology $\mT(\mV,\mE).$ The only task remaining in constructing the topology $\mT(\mV,\mE)$ is finding the {\color{black}true parents or children of the hidden nodes}. For this purpose, we consider each connected component $(M_l,Q_l),$  and reconstruct the topology associated with the hidden node $l$.

The following result is useful in reconstructing the exact topology associated with hidden node $l$ from \emph{degree} of nodes in $(M_l,Q_l)$. It shows that the information about the degree of each node is \emph{sufficient} to reconstruct the topology associated with the hidden node $l.$
\begin{theorem}
\label{theorem:two_Par_Sp}
{\color{black} Suppose the LDM in \eqref{eq:TF_block_model} satisfies Assumptions \ref{ass:atleastChild}-\ref{ass:diff_eqn}}.
Consider a hidden node $l\in \mV_h$ and its associated undirected graph $(M_l,Q_l)$ as defined in \black{Theorem \ref{thm2}}. Define $\alpha_l := max_{j \in M_l}^{} deg_{M_l}(j)$, where $deg_{M_l}(i):=|\{j\in M_l\setminus i: (i,j) \in Q_l\}|$.  The following holds: 
$\exists k \in M_l$ such that $deg_{M_l}(k)< \alpha_l$ if and only if $|\mathcal{P}(l)\setminus(\mathcal{C}(l)\cup \mathcal{S}(l) )|\geq 2$ or $|\mathcal{S}(l)\setminus(\mathcal{C}(l)\cup \mathcal{P}(l) )|\geq 2$ in $\mG(\mV,\mE).$
\end{theorem}
\hspace{-0.65cm}\begin{proof}
See supplementary material\black{, Appendix H} or \cite{doddi2019topology}.
\end{proof}

\begin{remark}\label{remark1}
From the proof of the above theorem, the following holds:
\begin{enumerate}[label=(\alph*)]
\item Consider a node $a_1\in \mathcal{C}(l)\cup (\mathcal{P}(l)\cap \mathcal{S}(l)).$ Then, for every $a_2 \in (\mathcal{C}(l) \cup \mathcal{P}(l)\cup \mathcal{S}(l))\setminus\{a_1\}, (a_1,a_2)\in Q_l.$ 
\item Consider a strict spouse, $s_1\in \mathcal{S}(l)\setminus (\mathcal{C}(l)\cup \mathcal{P}(l) )$. Then, for any $a_1 \in \mathcal{C}(l)\cup \mathcal{P}(l), (s_1,a_1)\in Q_l.$ 
\item Consider a strict parent, $p_1\in \mathcal{P}(l)\setminus (\mathcal{C}(l)\cup \mathcal{S}(l) )$. Then, for any $a_1 \in \mathcal{C}(l)\cup \mathcal{S}(l)\setminus\{p_1\}, (p_1,a_1)\in Q_l.$
\end{enumerate}


Regardless of the number of strict spouses and strict parents in the {\color{black} LDG}, $\mG(\mV,\mE),$ for $M_l = \mathcal{C}(l) \cup \mathcal{P}(l)\cup \mathcal{S}(l),$ the following holds from $(a),(b)$ and $(c)$: for $i \in \mathcal{C}(l)\cup(\mathcal{P}(l)\cap \mathcal{S}(l))$ and for $k \in M_l\setminus i, (i,k)\in Q_l.$ That is, a node from $\mathcal{C}(l)\cup(\mathcal{P}(l)\cap \mathcal{S}(l))$ is connected to every other node from $M_l$ in $Q_l.$ Hence, $deg_{M_l}(i)=|M_l|-1$ and it is the node with maximum degree. Therefore, $\mathcal{C}(l)\cup \{ \mathcal{P}(l)\cap \mathcal{S}(l) \}= \{i| i\in M_l, deg_{M_l}(i)=|M_l|-1 \}.$ The nodes in $M_l$ with $deg_{M_l}(i)< |M_l|-1$ are either strict parents or strict spouses of hidden node $l.$
\end{remark}

We provide the following assumption which is needed for exact reconstruction of topology associated with hidden nodes. When the below assumption is violated, that is, if the number of strict spouse for a hidden node $h$ is one, then there will be a single false edge present in the reconstructed topology associated with the hidden node. Rephrasing, the strict spouse of a hidden node $h$ will be considered as a neighbor in the reconstructed topology associated with the hidden node $h.$ The number of such false edges reconstructed for a network is limited to a maximum of one per hidden node. Nevertheless, to avoid these false edges associated with a hidden node, we make the following assumption.
\begin{assump} 
\label{ass:strict_spouse}
In the {\color{black} LDG}, $\mG(\mV,\mE)$, for any hidden node, $h \in \mV_h$, if there exists a strict spouse, then there must exist at least one more strict spouse associated with $h$. 
\end{assump}

Based on the above theorem and assumption, we propose Algorithm \ref{alg:top_hid}, which outputs the reconstructed topology $\mT(\mV_R,\mE_R)$ that is identical to the true topology $\mT(\mV,\mE)$ of the {\color{black} LDG} $\mG(\mV,\mE).$ The Algorithm $3$ consists of three parts as outlined below.
\begin{enumerate}[label=(\alph*)]
    \item From support of  $\Im\{\L\},$ determine $V_H$ and $E_H.$ The graph $(V_H,E_H)$ will be a disjoint collection of connected subgraphs $\bigcup\limits_{l=1}^{n_h} (M_l,Q_l).$ The number of hidden nodes, $n_h$, is given by the number of connected subgraphs. This is done in steps $1-10.$ $\mathcal{V}_H$ and $\mathcal{E}_H$ are both initialized with $\{ \}$ (steps $11-12$). 
    \item 
    For each $(M_l,Q_l),$ we create a hidden node $h_l$ and add this to $\mathcal{V}_H$ (steps $15-16$). Next, we construct the topology associated with $h_l.$ For this, we compute the degree of each node in $(M_l,Q_l)$ and calculate its maximum, $\alpha_l$ (step $17$). 
    We check if there is a node in $M_l$ with degree smaller than $\alpha_l.$ If there is no node in $M_l$ with degree smaller than $\alpha_l,$ then add undirected edge $(h_l,i)$ to $\mathcal{E}_H$ for all $i \in M_l$ (steps $18-28$). Otherwise, we collect the nodes in $M_l$ with degree $\alpha_l$ in the set $d_{h_l}$ and the nodes with degree smaller than $\alpha_l$ in the set $\widetilde{M}_l$ (steps $29-38$). The vertex set $d_{h_l}= \mathcal{C}(h_l)\cup \mathcal{P}(h_l)\cup \mathcal{S}(h_l).$ Thus, add $(h_l,i)$ to $\mathcal{E}_H$ for all $i \in d_{h_l}$ (steps $39-41$). The nodes in $\widetilde{M}_l$ are either strict parents of $h_l$ or strict spouses of $h_l$ in $\mG(\mV,\mE).$ We find the strict parents from $\widetilde{M}_l$ and add their edges with $h_l$ to $\mathcal{E}_H$ (steps $42-47$).
    \item Repeat (b) for all $l= \{1,2, \cdots, n_h \}$ (step $13$). Assign $\mV_R$ as $\mV_H\cup \mV_o$ and $\mE_R$ as $\mE_H\cup \mE_o$ (steps $49-50$). The reconstructed topology of the {\color{black} LDG} $\mG(\mV,\mE)$ is $\mT(\mV_R,\mE_R)$ (step $51$).  
\end{enumerate}

The reconstructed topology $\mT(\mV_R,\mE_R)$ is identical to the true topology $\mT(\mV,\mE)$ of the {\color{black} LDG} $\mG(\mV,\mE).$

{
{\tiny
\begin{algorithm}
\caption{Full Topology reconstruction with hidden nodes}
\textbf{Input:} $\hL(z)$ from Algorithm \ref{alg:matrix_decomposition},  threshold $\tau$ and $\mT(\mV_o,\mE_R)$ from Algorithm \ref{alg:top_rec}\\
\textbf{Output:} $\mT(\mV_R,\mE_R),$ reconstructed topology of the {\color{black} LDG} $\mG(\mV,\mE)$
\label{alg:top_hid}
\begin{algorithmic}[1]
    \State $V_H \gets \{ \}$
    \State $E_H \gets \{ \}$
	\ForAll{$(i,j) \in \{1,2,...,n\} \bigtimes \{1,2,...,n\}$}
	\If{$|\hL_{ij}(z)| >\tau$}
	\State $E_H \gets E_H \cup \{(i,j)\}$
	\State $V_H \gets V_H \cup \{i,j \}$
	\EndIf
	\EndFor
	\small
    \State $(V_H,E_H)= \bigcup\limits_{l=1}^{n_h} (M_l,Q_l)$ such that $M_{l_1}\cap M_{l_2} =Q_{l_1}\cap Q_{l_2}= \emptyset$ for all $l_1,l_2 \in \{1,2,...,n_h\}, l_1 \neq l_2.$
	\normalsize
	\State $n_h:$ number of disjoint connected subgraphs in $E_H$
	\State Vertex set ${\mathcal{V}}_H \gets \{ \}$  
	\State Edge set ${\mathcal{E}}_H \gets \{ \}$
  \ForAll{$l \in \{1,2,...,n_h\}$}
	\State flag = $0$
	\State add a hidden node $h_l$
	\State ${\mathcal{V}}_H \gets {\mathcal{V}}_H \cup \{h_l\}$
	\State $\alpha_l := max_{j \in M_l}^{} deg_{M_l}(j)$	
	\ForAll{$i \in M_l$}
	\If{$deg_{M_l}(i)< \alpha_l$}
    \State flag = $1$
    \State \textbf{break}
	\EndIf
	\EndFor
	\If{flag == $0$}
    \ForAll{$i \in M_l$}
	\State ${\mathcal{E}}_H \gets {\mathcal{E}}_H \cup \{(i,h_l)\}$    
    \EndFor
	\EndIf	
    \If{flag == $1$}
	
    \State $d_{h_l} \gets \{  \}$, $\widetilde{M}_l \gets \{  \}$
    \ForAll{$i \in M_l$}
        \If{$deg_{M_l}(i)== \alpha_l$}
    \State $d_{h_l} \gets \{d_{h_l}\} \cup \{i  \}$  
    \EndIf
    \If{$deg_{M_l}(i)< \alpha_l$}
    \State $\widetilde{M}_l \gets \widetilde{M}_l \cup \{i  \}$  
    \EndIf
    \EndFor

    \ForAll{$k \in d_{h_l}$}
	\State ${\mathcal{E}}_H \gets {\mathcal{E}}_H \cup \{(k,h_l)\}$    
    \EndFor
\ForAll{$k \in \widetilde{M}_l$}
\If{$(k,d)\notin \mathcal{E}_R$ for all $d \in d_{h_l}$}
\State ${\mathcal{E}}_H \gets {\mathcal{E}}_H \cup \{(k,h_l)\}$    
\EndIf
    \EndFor		
    \EndIf
	\EndFor
	
 	\State ${\mathcal{V}}_R \gets  \mathcal{V}_H \cup  \mathcal{V}_o $  
 	\State ${\mathcal{E}}_R \gets  \mathcal{E}_H \cup \mathcal{E}_R $  	\State Return $\mV_R$, $\mE_R$
\end{algorithmic}
\end{algorithm}
}
}

Till now, we have investigated topology identification under the assumption that the perfect PSDM is available. However, in practice, we have access only to finite number of observations of the time-series. In the next section, we show that if the number of observations, $N$, is large enough, then each entry of the estimated IPSDM can be brought $\epsilon$-close to the actual IPSDM. Note that we do not assume the presence of any latent node here. The analysis in this section is applicable to networks with or without latent nodes.
\section{IPSD Estimation from Finite time-series}
\label{sec:finite_TS}
In this section, we investigate the effect of finite time-series on the estimation of the IPSDM. {\color{black} Suppose that we are given the time-series $\{\x(t) \in \mathbb R^n\}_{t=1}^N$. We show that, with high probability, the entries of an estimated IPSDM are close to the original IPSDM, if the number of samples, $N$ is large enough}. To begin with, recall that \eqref{eqn:time_domain_model} can be rewritten as 
$\widetilde{\mathbf{x}}(t) = \sum_{k=-\infty}^{\infty} \widetilde{\bxi}(k) \boldsymbol{\widetilde{e}}(t-k), \ t \in \mathbb{Z}$,
where $\bxi(z):=(\I-\H(z))^{-1}$. Similar to \cite{goldenshluger2001}, for the convenience of analysis, we define: 

\vspace{-10pt}
{\footnotesize
\begin{align}
\label{set:H}
\mathcal{H}_\rho(l,L):=\{(\x(t), t\in \mathbb Z): 0<l\leq |\lambda(\bxi(z))|\leq L, \text{for } |z|\leq \rho \},
\end{align}
}where $\lambda(\bxi(z))$ denotes eigenvalues of $\bxi(z)$, and restrict ourselves to the family $\mathcal{H}_\rho(l,L)$.
\vspace{-10pt}
\subsection{Estimation Error}
Here, we characterize the difference between the actual PSDM, \black{$\Phi_{\x}(z)$, in \eqref{eq:psd}} and a PSDM estimated from finite time-series. We assume that the true correlation function $R_{\x}(k)=\mathbb{E}[\widetilde{\x}(0)\widetilde{\x}^T(k)]$ follows the relation 
\begin{align}
\label{eq:autocorrelation}
\linf{R_{\x}(k)}\leq C _1 \rho^{-k},
\end{align}
for some $\rho$ such that $|\rho| < 1$ (recall that $\widetilde{\x}$ is WSS).
Note that this assumption is common in scalar Gaussian processes and is an example of strongly mixing process (see \cite{goldenshluger2001} and the references therein for details). 

\black{Let $\widebar{\Phi}_{\x}(z)$ denote the truncated version of the PSDM, \black{$\Phi_{\x}(z)$,} and let $\widehat{\Phi}_{\x}(z)$ be the PSDM estimated from data. That is,
$\widebar{\Phi}_{\x}(z)=\sum_{k=-p}^p R_{\x}(k)z^{-k}$, and
$\widehat{\Phi}_{\x}(z)=\sum_{k=-p}^p \widehat{R}_{\x}(k)z^{-k}$,
where $\widehat{R}_{\x}(k)=\frac{1}{N-k}\sum_{l=1}^{N-k}\x(l)\x^T(l+k)$
is the estimated correlation matrix.}
We employ a concentration bound for $\linf{\Phi_{\x}(z)-\widehat{\Phi}_{\x}(z)}$ to bound $\linf{\Phi_{\x}^{-1}(z)-\widehat{\Phi}_{\x}^{-1}(z)}$\black{, for any $|z|=1$}. 
By triangle inequality,
{
\begin{align}
\nonumber
&\linf{\Phi_{\x}(z)-\widehat{\Phi}_{\x}(z)}\\ &\hspace{.5cm}\leq \linf{\Phi_{\x}(z)-\widebar{\Phi}_{\x}(z)}+\linf{\widebar{\Phi}_{\x}(z)-\widehat{\Phi}_{\x}(z)}.
\label{eq:finite_triangle}
\end{align}}
$\linf{\Phi_{\x}(z)-\widebar{\Phi}_{\x}(z)}$ is the truncation error, which is the error in truncating PSDM to order $p$, and $\linf{\widebar{\Phi}_{\x}(z)-\widehat{\Phi}_{\x}(z)}$ is the estimation error, which denotes the error in estimating the $p$-th order truncated PSDM.

The following proposition provides a bound on element-wise distance between $\widebar{\Phi}_{\x}$ and  ${\Phi}_{\x}$.
\begin{proposition}
\label{prop:PSD_diff_trunc} 
Consider a linear dynamic system governed by $\eqref{eqn:time_domain_model}$. Suppose that the autocorrelation function $R_{\x}(k)$ satisfies \eqref{eq:autocorrelation}.
For any $\eps>0$, if $p\geq\log_{\rho}\left( \frac{(1-\rho)\eps}{2C_1} \right)-1$, then the truncation error
$\linf{\Phi_{\x}(z)-\widebar{\Phi}_{\x}(z)} \leq \eps.$
\end{proposition}
\hspace{-0.65cm}\begin{proof}
See supplementary material\black{, Appendix D} or \cite{doddi2019topology}. 
\end{proof}

Note that
{\small
\begin{align}
\nonumber
\linf{\widebar{\Phi}_{\x}(z)-\widehat{\Phi}_{\x}(z)}&=\Linf{\sum_{k=-p}^{p} \left[R_{\x}(k)-\widehat{R}_{\x}(k)\right]z^{-k} }\\
\label{eq:finite_PSD_to_Rx_ineq}
&\leq \sum_{k=-p}^{p} \linf{R_{\x}(k)-\widehat{R}_{\x}(k)}.
\end{align}} 
Thus, by obtaining a bound on $\linf{R_{\x}(k)-\widehat{R}_{\x}(k)}$, we can upper bound $\linf{\widebar{\Phi}_{\x}(z)-\widehat{\Phi}_{\x}(z)}$.

Next, we provide probably approximately correct (PAC) bounds for estimating auto-correlation matrices, which in turn is used in obtaining PAC bounds on PSDMs.
The following proposition bounds deviation of each individual elements of auto-correlation matrix.
\begin{proposition}
\label{prop:linf_PAC_bound}
For every delay index $k\leq N-n$, $k \in \{0,\dots,p\}$, and all $\eps>0$, we have
{\small
\begin{align}
\nonumber
&\mathbb{P}\left( \linf{\widehat{R}_{\x}(k)-R_{\x}(k)}> \eps \right) \\
&\hspace{.5cm}\leq n^2 \exp\left(-(N-k)\min\left\{\frac{\eps^2}{32 n^2C_1^2},\frac{\eps}{8 nC_1}\right\} \right).
\end{align}
}
\hspace{-0.75cm}\begin{proof}
See supplementary material\black{, Appendix E} or \cite{doddi2019topology}. 
\end{proof}
\end{proposition}

Now, we are ready to obtain the  following lemma, which bounds the estimation error in \eqref{eq:finite_triangle} using  \eqref{eq:finite_PSD_to_Rx_ineq}.
\begin{lemma}
	\label{lem:finite_error_pac}
	For every \black{$z=e^{j\omega},~\omega \in (-\pi,\pi]$}, the error in estimating truncated PSDM $\widehat{\Phi}_{\x}(z)$ is 
$\linf{\widebar{\Phi}_{\x}(z)-\widehat{\Phi}_{\x}(z)}\leq \eps$ with probability at least 
\vspace{-8pt}

{\footnotesize \begin{align*}
	1- n^2 \exp\left(-(N-p)\min\left\{\frac{\eps^2}{32  (2p+1)^2 n^2C_1^2},\frac{\eps}{8  (2p+1) nC_1}\right\} \right).
	\end{align*}}
	
\hspace{-0.85cm}\begin{proof}See supplementary material\black{, Appendix F} or \cite{doddi2019topology}. 
\end{proof}
\end{lemma}


By combining the above lemmas, we obtain the following theorem.

\begin{theorem}
\label{thm:PSD_error_pac} 
Consider a linear dynamic system governed by $\eqref{eqn:time_domain_model}$. Suppose that the autocorrelation function $R_{\x}(k)$ satisfies \eqref{eq:autocorrelation}. Let $0<\eps_1<\eps$. Then, for any $p$ such that
$p\geq\log_{\rho}\left( \frac{(1-\rho)(\eps-\eps_1)}{2C_1} \right),
$
we have $\linf{{\Phi}_{\x}(z)-\widehat{\Phi}_{\x}(z)}\leq \eps$, with probability at least

\vspace{-8pt}
{\footnotesize \begin{align*}
1- n^2 \exp\left(-(N-p)\min\left\{\frac{\eps_1^2}{32  (2p+1) ^2n^2C_1^2},\frac{\eps_1}{8  (2p+1) nC_1}\right\} \right).
\end{align*}}
\end{theorem}

\hspace{-0.65cm}\begin{proof}
	The proof follows by combining Lemma \ref{lem:finite_error_pac} and Proposition \ref{prop:PSD_diff_trunc}.
\end{proof}

\vspace{-7pt}
{\color{black}
\subsection{IPSDM Estimation Error}}

Next, we provide a bound for the difference between the original and the estimated IPSDMs in terms of the difference between corresponding PSDMs. 

\begin{theorem}
\label{theorem:IPSD_error_pac} Consider a linear dynamic system governed by $\eqref{eqn:time_domain_model}$ satisfying the Assumption  \eqref{set:H}. Suppose that the autocorrelation function $R_{\x}(k)$ satisfies \eqref{eq:autocorrelation}.  Let $0<\eps_1<\eps$. Then, for any $p$ such that
$p\geq\log_{\rho}\left( \frac{(1-\rho)(\eps-\eps_1)}{2C_1} \right)-1,$
we have 
\black{
\vspace{-20pt}
{\footnotesize
\begin{align}
\linf{\Phi_{\x}^{-1}-\widehat{\Phi}_{\x}^{-1}}\leq \frac{ \sqrt{n}L^2 \sigma_{e_{[1]}}^2}{l^4 \sigma_{e_{[n_o]}}^4}\left(\frac{\sqrt{n}\eps}{l^2 \sigma_{e_{[n_o]}}^2-\sqrt{n}\eps}\right)
\end{align}}}
with probability at least 
\vspace{-8pt}
{\footnotesize	
\begin{align*}	1- n^2 \exp\left(-(N-p)\min\left\{\frac{\eps_1^2}{32  (2p+1) ^2\black{n^2C_1^2}},\frac{\eps_1}{8  (2p+1) \black{nC_1}}\right\} \right),	\end{align*}}
\black{where $\sigma_{e_{[1]}}$ and $\sigma_{e_{[n_o]}}$ denote the largest and the smallest eigenvalues of $\Phi_e(z)$ respectively.}
\end{theorem}

\hspace{-0.75cm}\begin{proof}
See supplementary material\black{, Appendix G} or  \cite{doddi2019topology}.\end{proof}


\black{It follows from Theorem \ref{theorem:IPSD_error_pac} that the element-wise distance between $\Phi_{\x}^{-1}$ and $\widehat{\Phi}_{\x}^{-1}$ can be made arbitrarily small with high probability, by picking $N$ large enough.}

\vspace{-10pt}
\section{Simulation results}
\label{sec:simulation}
In this section, we validate our algorithms with simulations. All the simulations are performed in Matlab; to solve the optimization problem for matrix decomposition, we use YALMIP \cite{yalmip} with SDPT3 \cite{sdpt3} solver.

{\color{black} For the simulations, we assume that we have access to the perfect IPSDM, $\Phi_{oo}^{-1}$.
From $\Im\{\Phi_{oo}\}$ alone, we aim to reconstruct the topology of the \black{LDM} which involves (a) recover the topology restricted to observed nodes given by $\mT(\mV_o,\mE_o),$ (b) determine the number of hidden nodes $n_h$ in {\color{black} LDG} $\mG(\mV,\mE)$, and (c) reconstruct topology associated with each hidden node.
\begin{remark}
The objective of our simulation is to demonstrate the working of Algorithm \ref{alg:matrix_decomposition} to decompose $\Im\{\Phi_{oo}^{-1}\}$ of this network into $\hS=\Im\{\S\}$ and $\hL=\Im\{\L\}$. Then, by applying algorithms \ref{alg:top_rec} and \ref{alg:top_hid} on $\hS$ and $\hL$, we can retrieve the original topology.
\end{remark}}


\begin{figure}
\includegraphics[trim=0 0 0 0,clip, width=.95\linewidth]{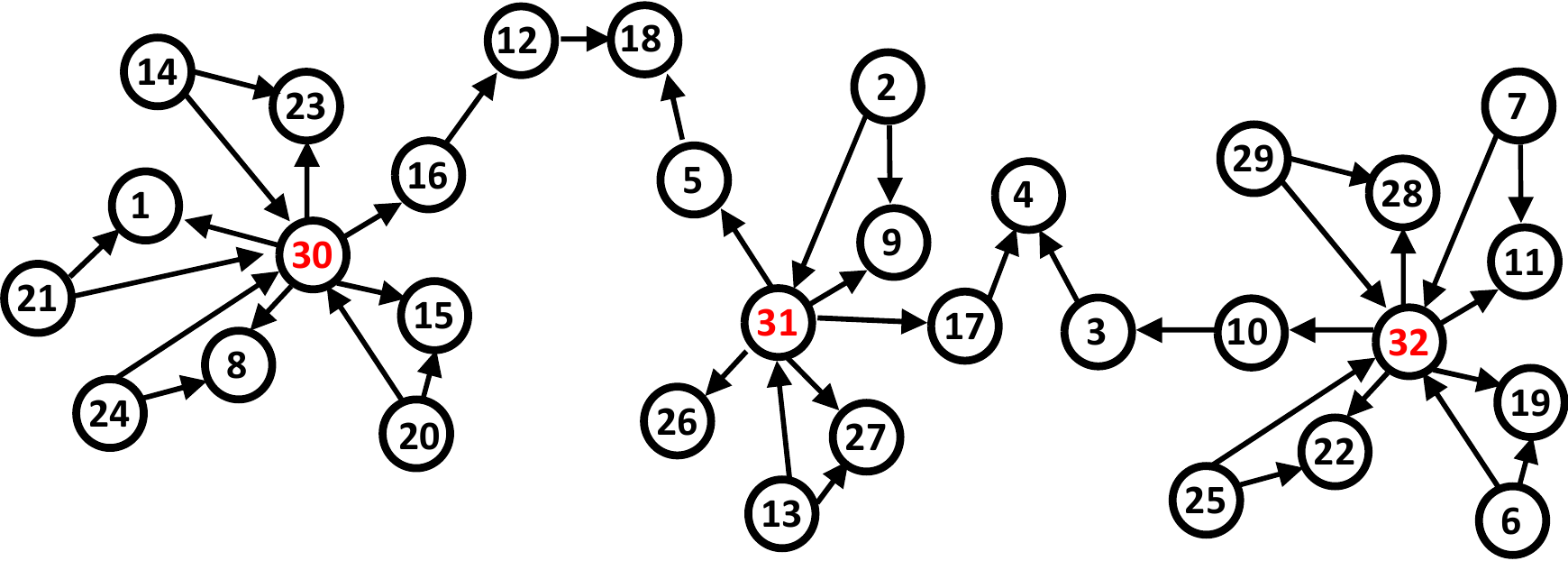}
\caption{{\color{black} LDG} $\mathcal{G}(\mV,\mE)$ with hidden nodes shown in red.}
\label{fig:gen_graph_new}
\end{figure}

\begin{figure}
	\centering
	\begin{tabular}{c}		\includegraphics[width=\linewidth]{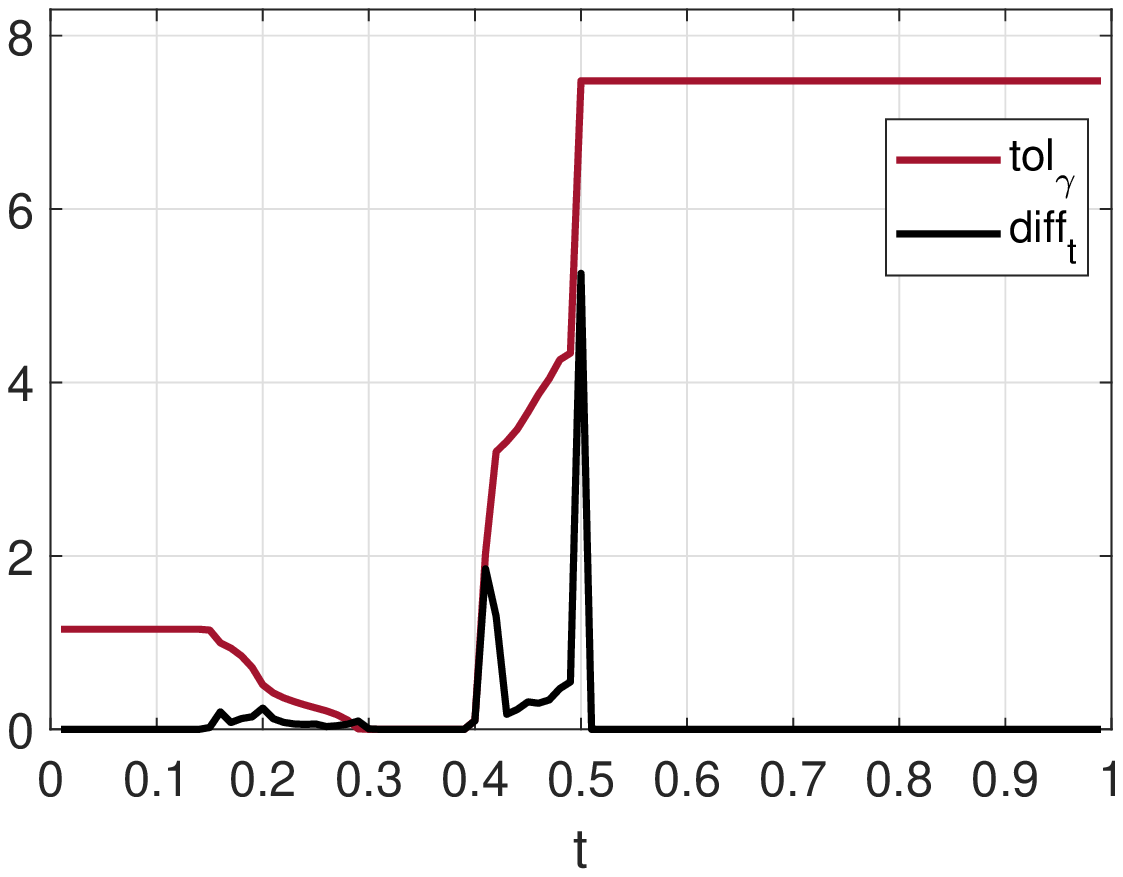}\\
	\end{tabular}
	\caption{Comparison between $tol_{t}$ and ${diff}_t$ for decomposition of $\Im\{\Phi_{oo}^{-1}\}$ obtained by the LDG in Fig. \ref{fig:gen_graph_new}}\label{fig:tol_diff_working}
\end{figure}

\vspace{-10pt}
\subsection{\black{LDM}}
{\color{black}For the simulation, we assume that $\Phi_{oo}$ of a topologically detectable LDM in \eqref{eq:TF_block_model} with $32$ nodes, having the LDG as shown in Fig. \ref{fig:gen_graph_new} is given. 
The nodes colored in red ($30,31,32$) are hidden and the rest are observed. Note that the network in Fig. \ref{fig:gen_graph_new} and the corresponding LDM satisfies the assumptions \ref{ass:atleastChild}-\ref{ass:strict_spouse}. 
} 

\begin{figure}
\centering
\begin{subfigure}[b]{0.95\textwidth}
\includegraphics[trim=0 0 0 0,clip, width=.5\textwidth]{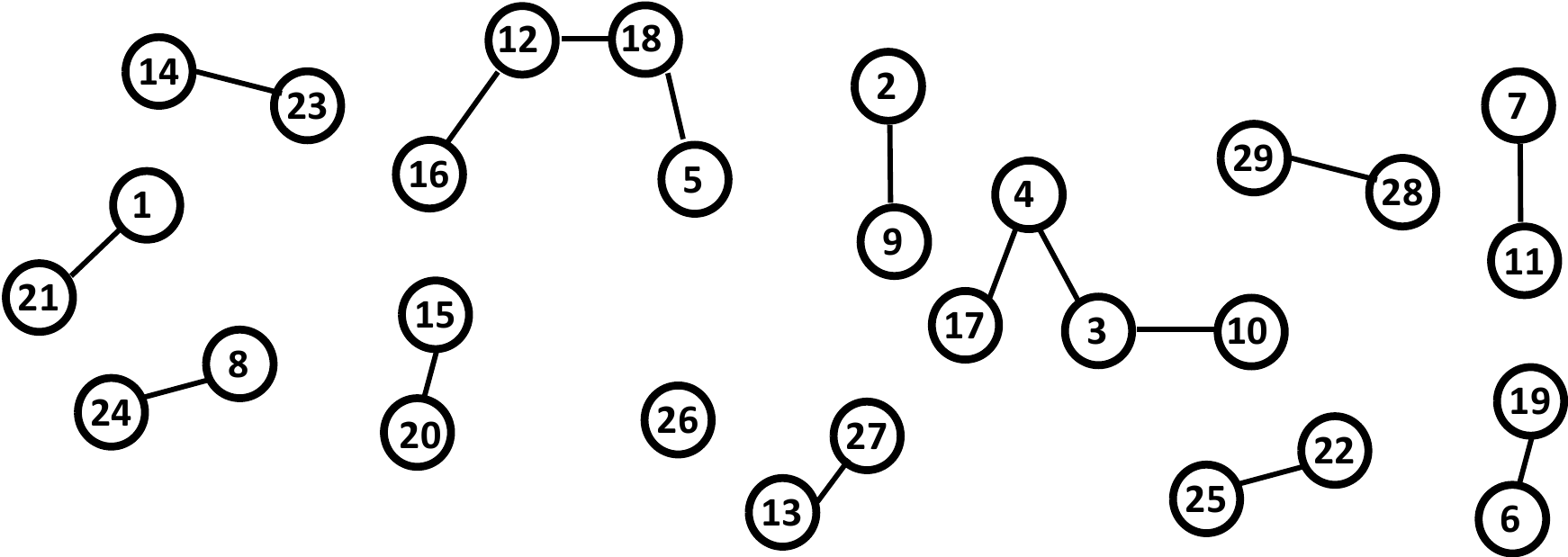}
\caption{Reconstructed topology among observable nodes, $\mT(\mathcal{V}_o,\mE_o),$\\ \hspace{.5cm}
obtained from $\Im\{\S\}$ using Algorithm $2$.}
\label{fig:top_observable}
\end{subfigure}

\begin{subfigure}[b]{0.95\textwidth}
\includegraphics[trim=0 0 0 0,clip, width=.5\textwidth]{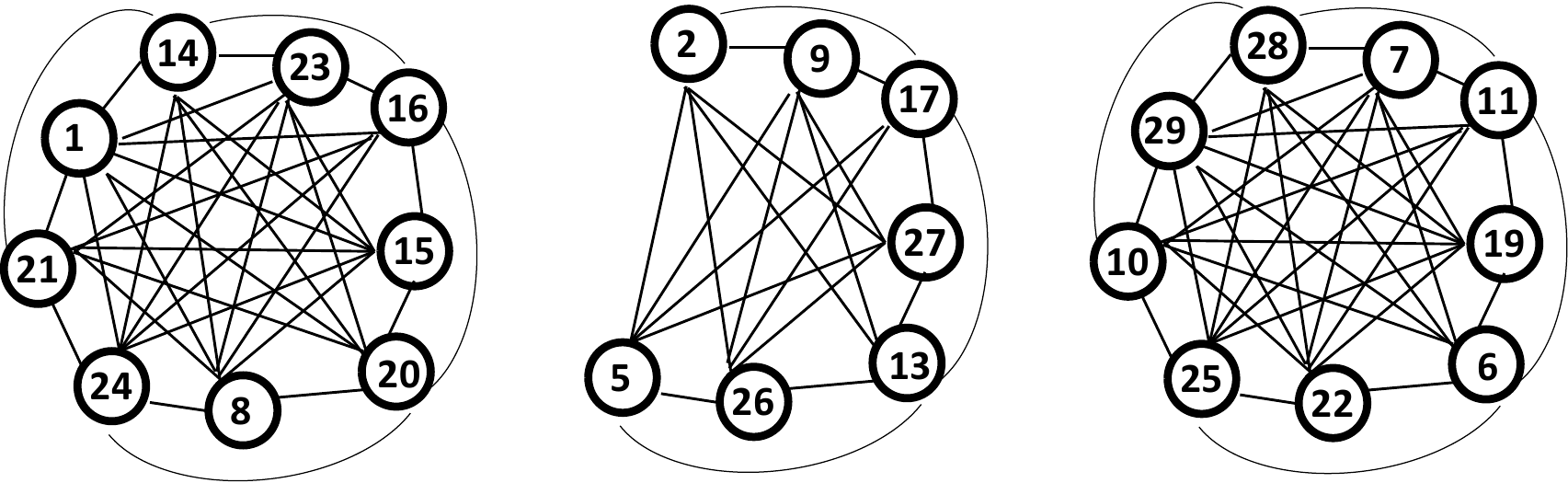}
\caption{ $(V_H,E_H)$ obtained from $\Im\{\L\}$ (step $9$ in Algorithm $3$).} 
\label{fig:top_hidden}
\end{subfigure}
\caption{Applying Algorithm $2$ on $\Im\{\S\}$ and Algorithm $3$ on $\Im\{\L\}$ using Algorithm $3$.} \end{figure}

\begin{figure}
\includegraphics[trim=0 0 0 0,clip, width=.95\linewidth]{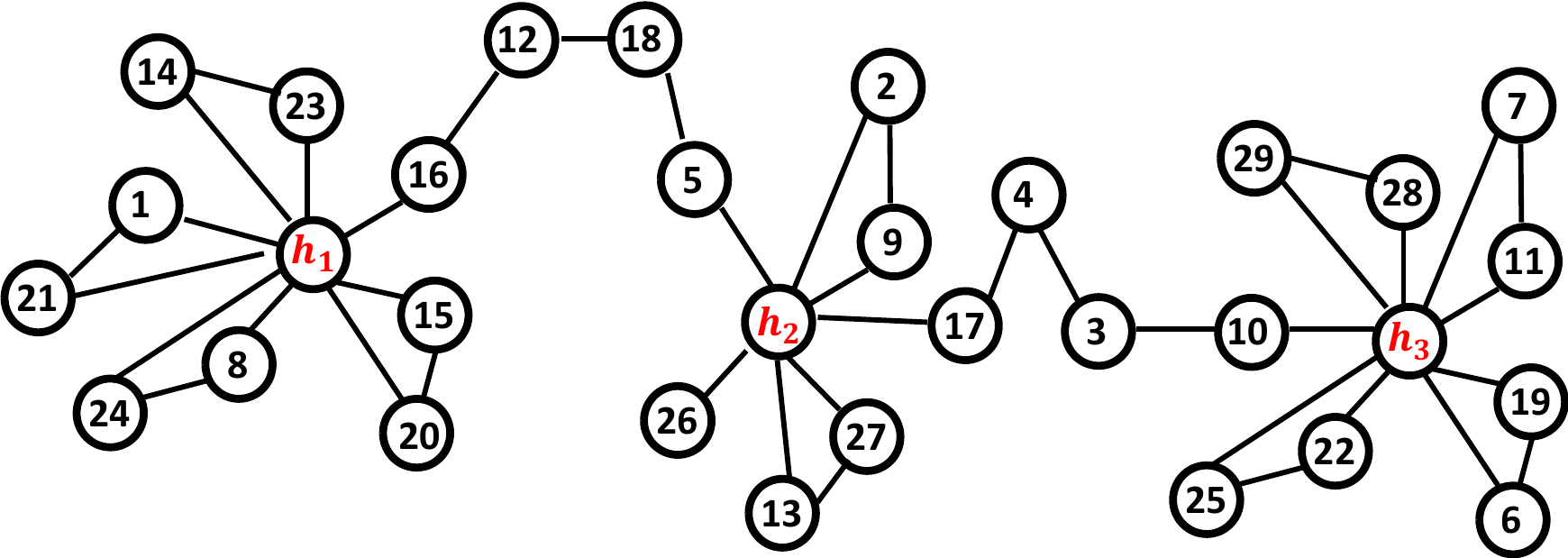}
\caption{$\mT(\mV_R,\mE_R)$: Final reconstructed topology with hidden nodes.}
\label{fig:new_top_reconstructed}
\end{figure}
\vspace{-10pt}
\subsection{Sparse plus low-rank decomposition of $\Im\{\Phi_{oo}^{-1}(z) \}$}
{\color{black}
 We applied Algorithm \ref{alg:matrix_decomposition} for matrix decomposition of $\Im\{\Phi_{oo}^{-1}(z)\}$ to obtain $\Im\{\S(z)\}=\hat{\S}$ and $\Im\{\L(z)\}=\hat{\L}$. Here, we took $z=e^{j\frac{3 \pi}{8}}$ and $\epsilon=0.01$ for running Algorithm \ref{alg:matrix_decomposition}.
Fig. \ref{fig:tol_diff_working} shows the values of $tol_{t}$ and ${diff}_t$ corresponding to the decomposition of $\Im\{\Phi_{oo}^{-1}(z)\}$ for various values of $t$.} From the plots, we see that the $diff_t$ plot shows three zero regions, similar to what was proposed in Proposition $\ref{prop:diff_t}$. We would like to stress that the network does not satisfy the sufficient condition of Lemma \ref{lem:suff_gamma_deg_inc}. Nevertheless, the plot can retrieve the true sparse and low-rank matrices by picking a $t$ in the middle zero region, as described by Algorithm \ref{alg:top_rec}. This network belongs to the subclass of the networks that does not satisfy the sufficient condition of Lemma \ref{lem:suff_gamma_deg_inc}, but satisfy the necessary and sufficient condition of \eqref{eq:transverse_intersection}. 

We make three important observations in Fig. \ref{fig:tol_diff_working}: (a) for small $t$, we observe a beginning zero region where ${diff}_t$ is zero. This verifies Proposition \ref{prop:diff_t}: when the value of $t$ is very small (smaller than $0.15$), the optimal objective value is obtained by setting zero to $\Im\{\L\}$ and the objective achieves the minimum value of zero at $(\Im\{\S\}+\Im\{\L\} ,0)$. As $t$ is increased but less than $0.15$, we still obtain the optimal solution at $(\Im\{\S\}+\Im\{\L\} ,0)$, and hence, ${diff}_t$ is zero for $t$ very small. 
(b) The opposite behavior is observed when $t$ is very large (greater than $0.5$). This is the end zero region, which corresponds to the optimal solution at $(0,\Im\{\S\}+\Im\{\L\} )$. Nevertheless, the value of $tol_{t}$ is quite high for both of these scenarios, since neither of the solutions are correct. (c) In the middle region ($t \in [0.24,0.35]$) , it is observed that both the $tol_{t}$ and ${diff}_t$ are zero; the range specified by Proposition \ref{prop:gamma_suff_muxi} (See Corollary \ref{cor:refResult1}).  

Note that $diff_t$ need not be zero in the middle region. However, it is observed from simulations that if we pick a $t$ corresponding to minimum value of $diff_t$ in the middle region between $(0.3, 0.4)$, it is still possible to reconstruct topology. This is because the finite value of $diff_t$ in middle region is due to element wise mismatch of non-zero entries. However, the structural pattern of sparse and low-rank matrices is almost preserved. For topology reconstruction, structural pattern is more important than the exactness of the element wise values. 


\vspace{-10pt}
\subsection{Reconstruction of $\mT(\mV,\mE)$}
Topology reconstruction of the {\color{black}LDM} involves (a) recover the topology restricted to observed nodes given by $\mT(\mV_o,\mE_o),$ (b) determine the number of hidden nodes $n_h$ in {\color{black} LDG} $\mG(\mV,\mE)$ and (c) reconstruct the topology associated with each hidden node.

From Fig. \ref{fig:tol_diff_working}, it is evident that $\Im\{\Phi_{oo}^{-1}(z)\}$ was decomposed uniquely into $\Im\{\S\}$ and $\Im\{\L\}$ {\color{black} when $t\in[0.3,0.4]$, since} $tol_{t}$ is zero. Indeed, due to perfect retrieval, the optimal solution is ($\Im\{\S\},\Im\{\L\}$). We now use the decomposed matrices $\Im\{\S\}$ and $\Im\{\L\}$ with $\tau=10^{-6}$ to obtain the following: 

\begin{enumerate}[label=(\alph*)]
\item From $\Im\{\S\}$, we apply Algorithm $2$ to obtain the topology of the subgraph restricted to the observable nodes. The reconstructed topology among observed nodes is shown in Fig. \ref{fig:top_observable}, which matches perfectly with topology of the {\color{black} LDG} in Fig. \ref{fig:gen_graph_new} restricted to observed nodes $\mT(\mV_o,\mE_o)$. 
\item From $\Im\{\L\},$ we apply Algorithm $3$ and construct ($V_H,E_H$), which is a union of three disjoint connected subgraphs, i.e., $(V_H,E_H)$=$\bigcup\limits_{l=1}^{3} (M_l,Q_l)$ (refer Fig. \ref{fig:top_hidden}). Therefore, number of hidden nodes in the {\color{black} LDG} are three hidden nodes present. $n_h$ is $3.$
    
    \item In Fig. \ref{fig:top_hidden}, each connected component $(M_l,Q_l)$ is a clique. For $l=\{1,2,3  \},$ Algorithm $3$ considers $l^{th}$ clique ($M_l,Q_l$), places a single hidden node $h_l$ in the clique and reconstructs the topology associated with hidden node $h_l.$ Since $(M_l,Q_l)$ is a clique, $M_l = \mathcal{C}(h_l)\cup \{\mathcal{P}(h_l)\cap \mathcal{S}(h_l) \}.$  
\end{enumerate}

From (a), (b), and (c), Algorithm $3$ reconstructs the complete topology including latent nodes (see Fig. \ref{fig:new_top_reconstructed}), which matches exactly with the topology of the {\color{black} LDG}. To summarize, Algorithm $1$ decomposed $\Im\{\Phi_{oo}^{-1}(z) \}$ into sum of sparse $\Im\{\S\}$ and low-rank $\Im\{\L\}.$ Using $\Im\{\S\},$ Algorithm $2$ reconstructs the topology reconstructed to observable nodes. Using $\Im\{\L\},$ Algorithm $3$ estimates the number of hidden nodes, reconstructs the topology associated with each hidden node and the full topology of the {\color{black} LDG} in Fig. \ref{fig:gen_graph_new} is reconstructed as shown in Fig. \ref{fig:new_top_reconstructed}, which matched exactly with true topology of {\color{black}the LDG}.
\vspace{-10pt}
\section{Conclusions}
\label{sec:conclusion}
{\color{black}We presented a novel approach to reconstruct the topology of networked linear dynamical systems with latent agents, from IPSDM of the observed nodes. The network was allowed to have directed loops and bi-directed edges. It was shown that the IPSDM can be uniquely decomposed into a sparse and a low-rank matrices. The sparse component unveiled the moral graph of the observed nodes, and the low-rank component retrieved the Markov Blanket associated with the latent nodes. Necessary and sufficient conditions for unique sparse plus low-rank decomposition of a skew symmetric matrix was established, along with an optimization based algorithm that decompose the skew symmetric matrix to yield the sparse component $\mathbf{S}$ and the low-rank component $\mathbf{L}$. For a large class of systems, the unique decomposition of imaginary part of the IPSDM is sufficient to achieve the moral graph of the observed nodes including the Markov Blanket of latent nodes. Under some assumptions, it was shown that the imaginary part of the IPSDM can be employed to reconstruct the exact topology of the network. Furthermore, concentration bounds on IPSDM estimation from finite time-series is provided. 
}
\appendix

\vspace{-5pt}
\subsection{Proof of Lemma \ref{lem:TS_sparse}}
\label{App:TS_sparse}
We show the result for an $n=3$ skew symmetric matrix. It is straight forward to extend this to a general $n$. Let $\A$={\small$\left[ \begin{array}{ccc}
     0 & a &b \\
    -a & 0 & 0\\
    -b & 0 & 0
\end{array}\right]$} be a skew symmetric matrix. $\A$ can defined as a point on the variety of skew symmetric matrices with support at most 4, i.e., $\mathcal{S}(4)=\mathbb{V}(\X_{11},\X_{22},\X_{33},\X_{12},\X_{12}+\X_{21},\X_{13}+\X_{31}, \X_{23}+\X_{32})$
$\cup \mathbb{V}(\X_{11},\X_{22},\X_{33},\X_{13},\X_{12}+\X_{21},\X_{13}+\X_{31}, \X_{23}+\X_{32})$
$\cup \mathbb{V}(\X_{11},\X_{22},\X_{33},\X_{23},\X_{12}+\X_{21},\X_{13}+\X_{31},\X_{23}+\X_{32})$.

Define $f_{ii}:=\X_{ii}, \forall 1 \leq i \leq n$, $f_{ij}:=\X_{ij}+\X_{ji}, \forall 1 \leq i<j \leq n$, and $g_{ij}:=\X_{ij},\forall 1 \leq i,j \leq n$. Note that $\A$ is a point on the variety defined by $f_{ii}=0 ,\ 1 \leq i \leq n, \ f_{12}=0, \ f_{13}=0$, $f_{23}=0,$ and $ g_{23}=0$, and is non-singular point (with respect to the variety). The tangent space of $\mathcal{S}(4)$ at $\A$, $T(\A)$, is (see Proposition 9.6.2 in \cite{IVA}):
{\scriptsize
\begin{align}
\label{eq:TS_variety_sparse}
    T(\A)=\mathbb{V}\left( (d_{\A}(f_{ii}))_{i=1}^3, d_{\A}(f_{12}), d_{\A}(f_{13}),d_{\A}(f_{23}), d_{\A}(g_{23}) \right),
\end{align}
}where $d_{\A}(f):=\sum_{i,j=1}^3 \frac{\partial f}{\partial x_{ij}}(\A)(\X_{ij}-\A_{ij})$. It can be shown that $d_{\A}(f_{ii})=\X_{ii}, \forall 1 \leq i \leq 3$, $d_{\A}(f_{ij})=\X_{ij}+\X_{ji},\forall 1 \leq i <j\leq 3$, and $d_{\A}(g_{23})=\X_{23}$.
Plugging this in \eqref{eq:TS_variety_sparse}, we obtain 
{\footnotesize
\begin{align}
T(\A)=\left\{\left[\begin{array}{ccc}
     0 & \X_{12} &\X_{13} \\
    -\X_{12} & 0 & 0\\
    -\X_{13} & 0 & 0
\end{array}\right] : \X_{12},\X_{13} \in \mathbb{R} \right\}.
\end{align}
}
Clearly, this is the space of all skew symmetric matrices with support subset of support of $\A$. Extending this analysis to general $n$ at any point $\A$ gives us the result.

\vspace{-8pt}
\subsection{Proof of Lemma \ref{lem:TS_rank}}
\label{app:TS_rank}
The following lemma is useful in proving this.
\begin{lemma}
\label{lem:non_singular_decomposition}
If $\M$ is a  skew symmetric $n \times n$ matrix of rank $r$ ($r$ even or $n>r$ odd), then there exists a non-singular matrix $\P$ such that $\M=\P\Q \P^T$,
where {\tiny$\Q=\left[\begin{array}{cc}
     \J &  \mathbf{0}\\
    \mathbf{0} & \mathbf{0}
\end{array}\right]$} \text{ and }
 {\tiny$\J=diag\left(\left[\begin{array}{ccc}
 0& 1 \\
 -1 & 0
\end{array} \right], \left[\begin{array}{ccc}
 0& 1 \\
 -1 & 0
\end{array} \right],\dots,\left[\begin{array}{ccc}
 0& 0 \\
 0 & 0
\end{array} \right]\right).$}\end{lemma}

Let $\mathcal{R}(r):=\{\X \in \mathbb{R}^{n \times n}: rank(\X)=r, \  \X+\X^T=0 \}$ be the space of skew symmetric matrices with rank $r$. By Youla decomposition (Lemma \ref{lem:non_singular_decomposition}), this is equivalent to 
$\mathcal{R}(r)=\{\P\Q\P^T:  \P \in \mathbb{R}^{n \times n} \text{ invertible} \}$.

 As shown in \cite{kozhasov2020minimality}, $\mathcal{R}(r)$ is a {\color{black} smooth} manifold. Let $\M$ be an element in $\mathcal{R}(r)$. Let $\g:(-1,1)\rightarrow\mathcal{R}(r)$ be a {\color{black} smooth} map such that $\g(t)=\P(t)\Q \P^T(t)$ with $\g(0) = \P(0)\Q\P^T(0)=\M$, where $\P(t)$ is invertible . Then,
 $\frac{d\g(t)}{dt}=\frac{d\P(t)}{dt}\Q \P^T(t)+ \P(t) \Q \frac{d\P^T(t)}{dt}$.
For $t=0$, this can be written as
    $\left.\frac{d\g(t)}{dt}\right|_{t=0}=\Delta \M + \M \Delta^T$,
where $ \Delta= \frac{d\P}{dt}(0)\P^{-1}(0)$. Define $W:=\{\Delta \M + \M \Delta^T: \Delta \in \mathbb{R}^{n \times n}\}$ and let $\Gamma_M:= \{\g'(0): \g$ is a {\color{black} smooth} map from (-1,1) to $\mathcal{R}(r)\}$ be the tangent space with respect to $\mathcal{R}(r)$ at $\M$. We claim that $\Gamma_M=W$. $\Gamma_M \subseteq W$ is obvious. To show the converse, let $w \in W$. Let $\Delta$ be such that $\Delta \M+\M\Delta^T =w$, and choose $\P(t):=\P(0)+t \Delta\P(0)$. Invertibility of $\P(t)$ follows from invertibility of $\P(0)$ and continuity of determinant. We need to show that this map belongs to the tangent space $\Gamma_M$. Note that $\left.\frac{d\P(t)}{dt}\right|_{t=0}= \Delta \P(0)$ and $\left.\frac{d\g(t)}{dt}\right|_{t=0}=\Delta \M + \M \Delta^T$, which proves the claim. Plugging in $\M^T=-\M$ and CSVD, $\M=\U\Sigma \Q\U^T$, in the definition of $W$, and substituting $\X=-\Delta\U\Sigma\Q$ completes the proof.

\vspace{-9pt}
\subsection{Proof of Proposition \ref{prop:diff_t}}
\label{appendix:diff_t}


For any matrix $\A\in \mathbb{R}^{n \times n}$,
$\frac{1}{k_1}\|\A\|_1\leq \|\A\|_*\leq k_2\|\A\|_1,$ where $k_1,k_2$ are real numbers. Note that $k_1$ and $k_2$ can be functions of $n$. The tightest bound available is $k_1=n$ and $k_2=1$ \cite{matrix_analysis}.

Define $f(\S,\L,t):=t \|\S\|_1+(1-t)\|\L\|_*$
. Clearly, the minimum value at $t=0$ is $f(\hS_0,\hL_0,\mathbf{0})=0$, and is obtained at the point $(\hS,\hL)=(\C,\mathbf{0})$.
Next, consider $t=\epsilon$ for some $\epsilon$ close to zero. 
The objective value at $(\S,\L)=(\C-\N,\N)$ for $t=\epsilon$, where $\N\neq \mathbf{0}$ is \begin{align}
f(\S,\L,\epsilon)=\epsilon \|\C-\N\|_1+(1-\epsilon)\|\N\|_*.
\end{align} 
Let the objective value for $t =\epsilon$ at the point $(\C,\mathbf{0})$ be $g(\epsilon):=f(\C,\mathbf{0},\epsilon)=\epsilon \|\C\|_1$. This can be rewritten as 
\begin{align}
\nonumber
g(\epsilon)&=\epsilon \|\C-\N+\N\|_1 \leq \epsilon \|\C-\N\|_1 +\epsilon\|\N\|_1\\
&\leq \epsilon \|\C-\N\|_1 +\epsilon k_1\|\N\|_*,
\end{align}
where the first inequality follows from the triangle inequality, and the second inequality follows since $\|\N\|_1 \leq k_1\|\N\|_*$.

Then, $f(\S,\L,\epsilon)-g(\epsilon)\geq (1- (k_1+1)\epsilon) \|\N\|_*$. For $0<\epsilon<\epsilon_L$, where $\epsilon_L=\frac{1}{k_1+1}$, we have that $f(\S,\L,\epsilon)-g(\epsilon)>0$, since $\|\N\|_*>0$. 
Thus, we have shown that for $t=\epsilon$ the value of the objective function at $(\C-\N,\N)$
is strictly  greater than that at $(\C,\mathbf{0})$. 
Moreover, this is true for all $\N\neq \mathbf{0}$. Therefore, we can conclude that for $0\leq t<\epsilon_L$, the optimal solution is $(\C,\mathbf{0})$.

Similarly, it is easy to see that the minimum value at $t=1$, $f(\hS_1,\hL_1,1)=0$, and is obtained at the point $(\S,\L)=(\mathbf{0},\C)$. Proceeding similar to the above analysis for $t=1$, with $(\S,\L)=(\N,\C-\N)$ and $g(\epsilon):=f(\mathbf{0},\C,\epsilon)=(1-\epsilon)\|C\|_*$, we can conclude that  $f(\S,\L,\epsilon)-g(\epsilon)\geq (1- (k_2+1)\epsilon) \|\N\|_1$. For $\epsilon_U<t\leq 1$, where $\epsilon_U=\frac{k_2}{k_2+1}$, we have that $f(\S,\L,\epsilon)-g(\epsilon)>0$, for every $\N\neq \mathbf{0}$. Thus, for $t>\epsilon_U$, the optimal solution is $(\mathbf{0},\C)$

Proposition \ref{prop:gamma_suff_muxi} and Lemma \ref{lem:suff_gamma_deg_inc} showed that (with the change of variable $t=\gamma/(1+\gamma)$), for every $t$ within the specified range, the optimal solution is $(\hS_t,\hL_t)=(\tS,\tL)$. That is, if $\mu \xi<1/6$ or  $\deg_{\max}({\tS}) inc({\tL})<1/12$, then we can find an interval $(t_1,t_2)\subset[0,1]$ with $0<t_1<t_2<1$ such that $(\tS,\tL)= (\hat{\S}_t,\hat{\L}_t)$ for any $t \in (t_1,t_2)$. This implies that ${diff}_t=0$ for $t \in [t_1+\eps,t_2-\eps]$ and sufficiently small interval $\eps$.
Therefore, there exists at least three zero regions for the plot of ${diff}_t$ versus $t$ if any of the sufficient conditions mentioned above is satisfied and $\eps$ is small enough.

{\color{black} 
\subsection{Proof of Theorem \ref{theorem:S_L_structure}}
\label{appendix:sparse_LR}
To prove the first part, let $\overline{\mE}_o$ be the set of undirected edges obtained by removing direction from $\mE_o$, without repetition. Then, by Proposition \ref{prop:S_top_obs_rec},  $\overline{\mE}_o \cup \mE_o^{ss} \supseteq \{(i,j): \S_{ij} \neq 0,~i<j\}$, i.e., $|support(\S)|\leq 2|\overline{\mE}_o|+2|\mE_o^{ss}|+n\leq 2|{\mE}_o|+2|\mE_o^{ss}|+n$, where the additional $n$ is due to the diagonal entries of $\S$. 

To show the upper bound on rank, let $\A=\Psi^{*}\Lambda^{-1}\Psi$. 
Notice that $rank(\A)\leq \min \{ rank(\Psi), rank(\Lambda) \} \leq n_h$. 
Similarly, $rank(\H_{ho}^{*}\Phi_{e_h}^{-1}\H_{ho})\leq n_h$. Then, $rank(L) \leq 2n_h$ from \eqref{eq:L_def}.
}
\vspace{-8pt}

{\color{black}
\subsection{Proof of Theorem \ref{theorem:S_top_obs_rec}}
\label{app:theorem:S_top_obs_rec}
From \eqref{eq:S_def}, observe that $\S$ is a Hermitian matrix, which is true if and only if $\Im\{\S\}$ is anti-symmetric, and thus the diagonal entries of $\Im\{\S\}$ are zeros. Lemma \ref{lem:Wiener_filter} showed that if $\S_{ij} \neq 0$, then $(i,j) \in kin(\mG_o)$. Combining this with Lemma \ref{lem:spurious_edges}, we get if $\Im(\S_{ij})  \neq 0$ then $i \in \mC(j)$ or $i \in \mP(j)$. That is, $\widehat{\mE}_o \subseteq \overline{\mE}_o$.



To prove the equality, we need to show that $\widehat{\mE}_o \supseteq \overline{\mE}_o$. Let $i\neq j$ and suppose $[\H_{oo}]_{ij}\neq 0$. By expanding \eqref{eq:S_def}, {\footnotesize $\S_{ij}=(\Phi_{e_o}^{-1})_{ij}-\sum_{k=1}^n (\H_{oo}^*)_{ik}(\Phi_{e_o}^{-1})_{kj}-\sum_{k=1}^n (\Phi_{e_o}^{-1})_{ik} {(\H_{oo})}_{kj}$ $+\sum_{k=1}^n\sum_{l=1}^n (\H^*)_{ik}(\Phi_{e_o}^{-1})_{kl}(\H_{oo})_{lj}$}. Clearly $\S_{ij} \neq 0$ since $[\H_{oo}]_{ij}\neq 0$, except for a few pathological cases that occur with Lebesgue measure zero. It follows by Assumption \ref{ass:H_real_const} that $\Im\{\S_{ij}\} \neq 0$, except over a set of Lebesgue measure zero.
}
\vspace{-8pt}
\subsection{Proof of Theorem \ref{thm1}}\label{appendix:thm1}
Recall expression \eqref{eq:L_def} for $\L$. By expanding each term in \eqref{eq:L_def} we obtain that if $\H_{hh}=0,$ then $\Lambda$ is diagonal, and $\Psi(k_h,j)$ can be written as

{\small\begin{align}
\nonumber
&\Psi(k_h,j)= \H_{oh}^*(j,k_h) \Phi^{-1}_{e_{o_j}}+\H_{ho}(k_h,j) \Phi^{-1}_{e_{h_{k_h}}}\\
&- \sum_{k=1}^{n_o} \Phi^{-1}_{e_{o_k}}\H_{oh}^*(k,k_h) \H_{oo}(k,j) 
\end{align}}
Then, we can conclude the following from the aforementioned equations.
\begin{itemize}
\item If the hidden nodes are at least two hops away, then $\Lambda$ is real and diagonal.
\item If any of $k_h\rightarrow j$ or $j \rightarrow k_h$ or $k_h\rightarrow k\leftarrow j$ exists for some $k \in \mV_o$, then $\Psi_{k_h,j}\neq 0$
\item \textcolor{black}{For $i \neq j,$} $\L_{ij}$ is given by the expression \eqref{L_ij} below.
\end{itemize}
\textcolor{black}{
\vspace{-10pt}
\tiny
\begin{align}\label{L_ij}
    &\L_{i,j} =\sum_{l=1}^{n_h}[\H_{ho}^{*}(l,i)]\Phi_{e_{h,l}}^{-1}\H_{ho}(l,j) - \sum_{l=1}^{n_h} d_l \nonumber \\
    &\left[\H_{oh}(i,l)\Phi_{e_{h,l}}^{-1} + [\H_{ho}^{*}(l,i)]\Phi_{e_{h,l}}^{-1} -\sum_{k=1}^{n_o} \Phi_{e_{o,k}}^{-1}\H_{oh}(k,l)[\H_{oo}^{*}(k,i)] \right]\nonumber\\
    & \left[\H_{oh}^{*}(j,l)\Phi_{e_{h,l}}^{-1} + [\H_{ho}(l,j)]\Phi_{e_{h,l}}^{-1} -\sum_{k=1}^{n_o} \Phi_{e_{o,k}}^{-1}\H_{oh}^{*}(k,l)[\H_{oo}(k,j)] \right]
\end{align}
\normalsize
}
Let $a_1$, $a_2$, and $a_3$ respectively denote first, second, and third terms in the first bracket and let $b_1$, $b_2$, and $b_3$ respectively denote first, second, and third terms in the second bracket. Then, (\ref{L_ij}) can be rewritten as,
\small
\begin{align} 
\nonumber
\L (i,j) &= \sum_{l=1}^{n_h}[\H_{ho}(l,i)]^{*}\Phi_{e_{h,l}}^{-1}\H_{ho}(l,j) -\\
\nonumber
&\sum_{l=1}^{n_h} d_l\left[a_1+a_2+a_3 \right]\left[b_1+b_2+b_3\right]\\
\nonumber
=&\ \sum_{l=1}^{n_h}[\H_{ho}(l,i)]^{*}\Phi_{e_{h,l}}^{-1}\H_{ho}(l,j) -\\
\nonumber
&- \sum_{l=1}^{n_h} d_l[a_1b_1+a_1b_2+a_1b_3+ a_2b_1+a_2b_2+\\
\label{eq:Lij_withoutsimplif}
& \ \ \  \ \ \  \ \ \ a_2b_3+ a_3b_1+a_3b_2+a_3b_3].
\end{align}
\normalsize

\begin{enumerate}[label=(\alph*)]
\item We use contrapositive argument to prove this. Suppose that, for every hidden node $h\in \mV_h,$ there does not exist $g \in DM_h(i,j)$ such that $g \in \mG(\mV,\mE).$ Then, from \textcolor{black}{(\ref{eq:Lij_withoutsimplif})}, it follows that  \textcolor{black}{${\L}_{ij} = 0.$} Therefore, if \textcolor{black}{${\L}_{ij} \neq 0,$} then there exists $g \in DM_h(i,j)$ such that $g \in \mG(\mV,\mE)$, for some $h\in \mV_h$. Clearly, for any $g \in DM_h(i,j),$ we have {\color{black}$\dh({i,h})\leq 2$ and $\dh({j,h})\leq  2.$}
\item Let $i,j\in \mV_o$. Suppose that \textcolor{black}{${\L_{ij}}\neq 0.$} Then, from part (a), 
there exists $g_1 \in DM_{h_1}(i,j)$ such that $g_1 \in \mG(\mV,\mE)$, for some $h_1\in \mV_h$. Suppose for contradiction that there exists $h_2 \in \mV_h\setminus \{h_1\}$ such that there exists $g_2 \in DM_{h_2}(i,j)$ such that $g_2 \in \mG(\mV,\mE).$ {\color{black}Then, we have $\dh({h_1,h_2})\leq \dh({h_1,i})+\dh({i,h_2}) \leq 4$ from Theorem \ref{thm1}(a)},
which contradicts Assumption \ref{ass:latent5hop}. Hence, $h_1= h_2.$ 



\end{enumerate}

\vspace{-8pt}
\subsection{Proof of Theorem \ref{thm2}}\label{appendix:thm2}
\begin{enumerate}[label=(\alph*)]
\item Suppose $M_{l_1}\cap M_{l_2}\neq \emptyset$ for some $l_1,l_2 \in \mV_h, l_1\neq  l_2$. Then, there exists $a\in M_{l_1}\cap M_{l_2}.$ Since $a \in M_{l_1},$ it follows from the definition of $M_{l_1}$ that $\dh({a,l_1})\leq 2.$ Similarly, $a \in M_{l_2}$ and $\dh({a,l_2})\leq 2$.  Then, we have
$\dh({l_1,l_2})\leq \dh({l_1,a})+\dh({a,l_2})\leq 4$, which contradicts Assumption \ref{ass:latent5hop}. Therefore, $M_{l_1}\cap M_{l_2}=\emptyset$ for all $l_1,l_2 \in \mV_h, l_1\neq l_2.$
\item Suppose $Q_{l_1}\cap Q_{l_2}\neq\emptyset$ for some $l_1,l_2 \in \mV_h, l_1\neq l_2$. Then, there exists $(i_0,j_0)\in Q_{l_1}\cap Q_{l_2}.$ $(i_0,j_0)\in Q_{l_1}$, which implies that $i_0,j_0 \in M_{l_1}.$ Similarly, $(i_0,j_0)\in Q_{l_2}$ implies that $i_0,j_0 \in M_{l_2}.$ Thus, $i_0,j_0 \in M_{l_1} \cap M_{l_2}$, which contradicts part $(a)$. Therefore, $Q_{l_1}\cap Q_{l_2}=\emptyset.$ 
\item We first show that $V_H \supseteq \bigcup\limits_{l=1}^{n_h} M_l$. To show this, let $a \in \bigcup\limits_{l=1}^{n_h} M_l$. Then, $a \in M_l$ for some $l \in \mV_h.$ It follows from the definition of $M_l$ that $a \in \mathcal{P}(l)\cup \mathcal{C}(l) \cup \mathcal{S}(l).$ If $a \in \mathcal{P}(l)\cup \mathcal{C}(l),$ then by Assumption \ref{ass:atleastChild}, there exists $c\in M_l\setminus{a}.$ If $a \in \mathcal{S}(l)\setminus (\mathcal{C}(l)\cup \mathcal{P}(l)),$ then there exists $k \in M_l\setminus{a},$ such that $k\in \mathcal{C}(l)\cap \mathcal{C}(a)$. Given the existence of node $k\in \mathcal{C}(l),$ by Assumption \ref{ass:atleastChild} there exists $c\in M_l\setminus \{a,k\}.$ Clearly there is a node $c\in M_l\setminus a$ in $\mG(\mV,\mE),$ which is also present in $DM_l(a,c).$ From \textcolor{black}{(\ref{eq:Lij_withoutsimplif})} and Remark \ref{rem:pathological}, it follows that \textcolor{black}{${\L}_{ac} \neq0.$} Because \textcolor{black}{${\L}_{ac} \neq0, $} it follows from the definition $V_H,$ that $a\in V_H.$

Now, we show that $V_H \subseteq \bigcup\limits_{l=1}^{n_h} M_l$. Let $a \in V_H,$ then from the definition of $V_H,$ there exists $c\in \mV_o\setminus{a},$ such that \textcolor{black}{${\L}_{ac} \neq0.$} From Theorem \ref{thm1} parts (a) and (b), there exists a unique hidden node $l \in \mV_h$ such that $g \in DM_l(a,k)$ exists in $\mG(\mV,\mE).$ From the definition of $DM_l(a,k),$ it follows that $a\in M_l.$ Hence, $a \in \bigcup\limits_{l=1}^{n_h} M_l$, which concludes the proof.

We first show $E_H \supseteq \bigcup\limits_{l=1}^{n_h} Q_l$. Let $(i_0,j_0) \in \bigcup\limits_{l=1}^{n_h} Q_l, $ then there exists a $l \in \mV_h,$ such that $(i_0,j_0) \in Q_l.$ From the definition of $Q_l, \{i_0,j_0\} \in M_l \subset V_H$ and \textcolor{black}{${\L}_{i_0 j_0} \neq0.$} Thus $(i_0,j_0) \in E_H.$
    
To show the converse $E_H \subseteq \bigcup\limits_{l=1}^{n_h} Q_l$, let $(i_0,j_0) \in E_H,$ then \textcolor{black}{${\L}_{i_0 j_0} \neq 0.$} From Theorem \ref{thm1}(b), there exists a unique hidden node $l\in \mV_h,$ such that a $g \in DM_l(i_0,j_0)$ exists in $\mG(\mV,\mE).$ It follows from the definition of $DM_l(i_0,j_0)$ that $\{i_0,j_0 \}\in M_l.$ $\{i_0,j_0 \}\in M_l.$ and \textcolor{black}{${\L}_{i_0 j_0} \neq 0,$} thus $(i_0,j_0)\in Q_l \subset \bigcup\limits_{l=1}^{n_h} Q_l.$
\end{enumerate}
\vspace{-8pt}

	\nocite{bruce}

\begin{figure*}[t!]
    \centering
\textbf{\Huge{Supplementary Material: Topology Learning of Linear Dynamical Systems with Latent Nodes using Matrix Decomposition}}
\end{figure*}

\newpage
\newpage

\import{./}{Supplementary.tex}

\end{document}

%% file: Supplementary.tex
\newpage

\appendix

\subsection{Proof of Lemma 3:}

SVD of a matrix $\M$ is given by $\M=\U\Sigma \V^T$, where $\U,\V \in \mathbb{R}^{n \times r} $ are left and right singular matrices and $\Sigma=diag(\lambda_1,\lambda_1,\dots,\lambda_{r/2},\lambda_{r/2})$.
By Youla decomposition, any skew symmetric matrix $\M$ of rank $r$ can be written in the form $\M=\W\Q\W^T$, where 
{\footnotesize
\begin{align*}
\Q= \begin{array}{ccc}
diag\left(\left[\begin{array}{ccc}
0& \lambda_1 \\
\lambda_1 & 0
\end{array} \right],\dots,\left[\begin{array}{ccc}0& \lambda_{r/2} \\
\lambda_{r/2} & 0
\end{array} \right],\left[\begin{array}{ccc}
 0& 0 \\
 0 & 0
\end{array} \right]\dots,\right)
\end{array}
\end{align*}}and $\W$ is an orthogonal matrix. Note that this is equivalent to writing $\M=\W_{r} \Sigma \widetilde{\Q} \W_{r}^T$, where $ \W_{r}$ is obtained by picking first $r$ columns of $\W$, $\Sigma=diag(\lambda_1,\lambda_1,\dots,\lambda_{r/2},\lambda_{r/2})$, and $\widetilde{\Q}=diag(\underbrace{\L,\dots,\L}_{r\text{ times}})$ with $\L=\left[\begin{array}{ccc}
 0& 1 \\
 -1 & 0
\end{array} \right]$. That is, the svd, $\M=\U\Sigma\V^T$, can be rewritten with $\V=\U\widetilde{\Q}^T$, where $\widetilde{\Q}^T=\widetilde{\Q}^{-1}$. 

Thus, $\V\V^T=\U\widetilde{\Q}^T\widetilde{\Q} \U^T=\U\widetilde{\Q}^{-1}\widetilde{\Q} \U^T=\U \U^T$. This concludes the proof.

\subsection{Proof of Proposition 1:}
\label{app:lagrangian_suff}

Subdifferential of a convex function $f$ at a point $\Y$ in the domain $\mathcal{X}$ is defined as \cite{boyd_vandenberghe_2004}:
{\small \begin{align}
\label{eq:subdiff_def}
\partial f(\Y)=\{g: f(\X) \geq f(\Y)+\langle g, \X-\Y \> , \forall \X \in \mathcal{X}\},
\end{align}}
where $\<\A,\B\>:=trace\{\A^T\B\}=\sum_{i,j} \A_{ij}\B_{ij}$.

The lagrangian of (13), $J$, can be written as 
\begin{align}
\nonumber
\label{eq:lagrangian}
    J(\S,\L,\Q_1,\Q_2)&=\g \|\S\|_1+\|\L\|_* + \<\Q_1,\C-\S-\L \>\\
    &+\<\Q_2,\S+\S^T \>
\end{align}
\begin{remark}
We can add another dual variable corresponding to the constraint $\L+\L^T=\mathbf{0}$. However, this would be redundant with $\Q_2$ as $\C$ is always a skew symmetric matrix.
\end{remark}
The optimality conditions are then given by 
\begin{align}
\Q_1-\Q_2-\Q_2^T  \in \gamma \partial \|\tS\|_1, \text{ and } \Q_1\in \gamma \partial \|\tL\|_*.
\end{align}
From the characterization of subdifferential of $\|\tS\|_1$, it follows that 
\begin{align}
\nonumber
&P_{\Omega(\tS)}(\Q_1-\Q_2-\Q_2^T ) = \gamma sign(\tS) \text{ and }\\ 
\label{eq:subdiff_l1}
&\hspace{1cm}\|P_{\Omega(\tS)^C}(\Q_1-\Q_2-\Q_2^T )\|_{\infty}\leq  \gamma,
\end{align}
where $P_{\Omega(\tS)}(\A)$ is obtained by setting entries of $\A$ outside the support of $\tS$ to zero and projecting it to the space of skew symmetric matrices.

Similarly, from the characterization of $\partial\|\tL\|_*$\cite{CSPW_siam_2011}, we have
\begin{align}
\label{eq:subdiff_nuclear}
&P_{T(\tL)}(\Q_1 ) = \U\V^T \text{ and } \|P_{T^{\perp}}(\Q_1)\|_{2}\leq  1,
\end{align}
where $P_{T(\tL)}(\A ):=P_{\U}\A+\A P_{\U}-P_{\U}\A P_{\U}$; $P_{\U}=\U\U^T$, $\tL=\U\Sigma\V^T$. 

From the subgradient optimality conditions, we have that $(\tS,\tL)$ is an optimal solution if there exist duals $\Q_1,\Q_2$ such that \begin{align}
\Q_1-\Q_2-\Q_2^T  \in \gamma \partial \|\tS\|_1, \text{ and } \Q_1\in \gamma \partial \|\tL\|_*.
\end{align}
The second condition in the proposition statement guarantees the existence of such a dual since they satisfy \eqref{eq:subdiff_l1} and \eqref{eq:subdiff_nuclear}. 

Next, we show the uniqueness of the solution. We prove this by contradiction. Let $(\Q_S,\Q_L)$ be any subdifferential at $(\tS,\tL)$ and let $ (\tS+\N_S,\tL+\N_L)$ be another optimal solution. It follows, from the constraint $C=\tS+\N_S+\tL+\N_L$, that $\N_S+\N_L=\zero$. Then,

{\footnotesize
\begin{align*}
\g \|\tS+\N_S\|_1+\|\tL+\N_L\|_* \geq \g \|\tS\|_1+\|\tL\|_* + \<\Q_{S},\N_S\>+\<\Q_{L},\N_L\>.
\end{align*}}
For notational simplicity, we write $\Omega(\tS)$, $\Omega^C(\tS)$, $T(\tL)$, and $T^{\perp}(\tL)$ as $\Omega$, $\Omega^C$, $T$ and $T^{\perp}$ respectively.
{\small\begin{align*}
&\< \Q_S, \N_S\>=\<\g sign(\tS)+P_{\Omega^C}(\Q_S),\N_S\>\\
&=\<\Q_1-\Q_2-\Q_2^T-P_{\Omega^C}(\Q_1-\Q_2-\Q_2^T)+P_{\Omega^C}(\Q_S),\N_S\>\\
\nonumber
&=\<\Q_1,\N_S\>- \<\Q_2+\Q_2^T,\N_S\>\\
&\hspace{1cm}\<-P_{\Omega^C}(\Q_1-\Q_2-\Q_2^T)+P_{\Omega^C}(\Q_S),\N_S\>\\
&\stackrel{(i)}{=}\<\Q_1,\N_S\>+\<-P_{\Omega^C}(\Q_1-\Q_2-\Q_2^T)+P_{\Omega^C}(\Q_S),\N_S\>.
\end{align*}}
Here, $(i)$ follows since $\Q_2+\Q_2^T$ is a symmetric matrix and $\N_S$ is skew symmetric matrix. Then,  $\<\Q_2+\Q_2^T,\N_S\>=0$. Similarly,
{\small\begin{align*}
\< \Q_L, \N_L\>&=\<\U \V^T + P_{T^{\perp}}(\Q_L),\N_L\>\\
&\stackrel{(ii)}{=}\<\Q_1-P_{T^{\perp}}(\Q_1)+P_{T^{\perp}}(\Q_L),\N_L\>\\
\nonumber
&=\<\Q_1,\N_L\>+\<P_{T^{\perp}}(\Q_L)-P_{T^{\perp}}(\Q_1),\N_L\>,
\end{align*}}
where $(ii)$ follows since $ \Q_1=\U\V^T+P_{T^{\perp}}(\Q_1)$.

Then,
{\footnotesize
\begin{align*}
&\< \Q_L, \N_L\>+\< \Q_S, \N_S\>=\<\Q_1,\N_L\>+\<P_{T^{\perp}}(\Q_L)-P_{T^{\perp}}(\Q_1),\N_L\>\\
&\hspace{1cm}+\<\Q_1,\N_S\>+\<P_{\Omega^C}(\Q_S)-P_{\Omega^C}(\Q_1-\Q_2-\Q_2^T),\N_S\>\\
&\hspace{1cm}\stackrel{(iii)}{=}\<P_{T^{\perp}}(\Q_L)-P_{T^{\perp}}(\Q_1),\N_L\>\\
&\hspace{2cm}+\<P_{\Omega^C}(\Q_S)-P_{\Omega^C}(\Q_1-\Q_2-\Q_2^T),\N_S\>\\
&\hspace{1cm}\stackrel{}{=}\<P_{T^{\perp}}(\Q_L)-P_{T^{\perp}}(\Q_1),P_{T^{\perp}}(\N_L)\>\\
&\hspace{2cm}+\<P_{\Omega^C}(\Q_S)-P_{\Omega^C}(\Q_1-\Q_2-\Q_2^T),P_{\Omega^C}(\N_S)\>.
\end{align*}
}
Here, $(iii)$ follows since $\N_S+\N_L=0$.
Now, choose $\Q_S=\g sign(P_{\Omega^C}(\N_S))$ and $\Q_L=\widehat{\U}\widehat{\V}^T$, where $P_{T^{\perp}}(\N_L)= \widehat{\U}\widehat{\Sigma} \widehat{\V}^T$. Applying this substitution along with  Holder's inequality \cite{matrix_analysis}, $\< \Q_L, \N_L\>+\< \Q_S, \N_S\>$ can be written as 
{\footnotesize
\begin{align*}
&\< \Q_L, \N_L\>+\< \Q_S, \N_S\> \geq\left(1-\|P_{T^{\perp}}(\Q_1)\|_2\right) \|P_{T^{\perp}}(\N_L)\|_*\\
&\hspace{1cm}+(\g-\| P_{\Omega^C}(\Q_1-\Q_2-\Q_2^T) \|_{\infty})\|P_{\Omega^C}(\N_S)\|_1.
\end{align*}}
Since $\| P_{\Omega^C}(\Q_1-\Q_2-\Q_2^T) \|_{\infty}<\g$ and $\|P_{T^{\perp}}(\Q_1)\|_2<1$, $\< \Q_L, \N_L\>+\< \Q_S, \N_S\>=0$ if and only if both $P_{\Omega^C}(\N_S)=P_{T^{\perp}}(\N_L)=\zero$, which implies that $P_{\Omega}(\N_S)+P_{T}(\N_L)=\zero$ as $\N_S+\N_L=\zero$. Since $\Omega(
\tS) \cap T(\tL)=\{\mathbf{0}\}$, this would mean that $\N_S=\N_L=\zero$.

\subsection{Proof of Proposition 2:}
It can be checked that if $\mu(\tS)\xi(\tL)<\frac{1}{6}$, then  
$\frac{\xi(\tL)}{1-4\mu(\tS)\xi(\tL)}<\frac{1-3\mu(\tS)\xi(\tL)}{\mu(\tS)}$ and the range in (22) is non-empty.

Here, we construct the duals $\widehat{\Q}_1,\widehat{\Q}_2$ by picking both from the direct sum $\Omega \oplus T$, that is, $\widehat{\Q}_1=\widehat{\Q}_{1,\Omega}+\widehat{\Q}_{1,T}$ and $\widehat{\Q}_2=\widehat{\Q}_{2,\Omega}+\widehat{\Q}_{2,T}$, where $\widehat{\Q}_{1,\Omega},\widehat{\Q}_{2,\Omega} \in \Omega$ and $\widehat{\Q}_{1,T},\widehat{\Q}_{2,T} \in T$. We will show that,
for any $\g$ taken from (22), the duals satisfy the sufficient conditions specified by Proposition 1. This would then guarantee that (13) returns $(\tS,\tL)$ as the unique optimum solution.

Let $\widehat{\Q}=\widehat{\Q}_1-\widehat{\Q}_2-\widehat{\Q}_2^T$. $\widehat{\Q}_2$ is a skew symmetric matrix by construction, and hence, $\widehat{\Q}_2+\widehat{\Q}_2^T=0$, which implies $\widehat{\Q}=\widehat{\Q}_1$. By following the similar arguments from Proposition $1$ in \cite{CSPW_siam_2011}, it can be shown that $\Omega \cap T=\{\zero\}$ if $\mu(\tS)\xi(\tL)<1/6$. It follows that we can find a unique $\widehat{\Q}_1$ such that $P_{\Omega}(\widehat{\Q}_1)=\g sign(\tS)$ and $P_{T}(\widehat{\Q}_1)=\U\V^T$. We will show that this particular $\widehat{\Q}_1$ satisfies the remaining conditions in Proposition 1.

Let $\widehat{\Q}_{1,\Omega}=\g sign(\tS)+\epsilon_{\Omega}$ and $\widehat{\Q}_{1,T}=\U\V^T+\epsilon_{T}$. Then, \begin{align}
P_{\Omega}(\widehat{\Q})&=P_{\Omega}(\widehat{\Q}_{1,\Omega})+P_{\Omega}(\widehat{\Q}_{1,T})\\
&\hspace{1cm}=\g sign(\tS) +\epsilon_{\Omega}+P_{\Omega}(\U\V^T+\epsilon_{T}).
\end{align}
From \eqref{eq:subdiff_l1}, it follows that $P_{\Omega}(\widehat{\Q})=\g sign(\tS)$, which indicates that 
$\epsilon_{\Omega}=-P_{\Omega}(\U\V^T+\epsilon_{T})$. Similarly, from the definition of $P_{T^{\perp}}(\widehat{\Q}_1)$ and \eqref{eq:subdiff_nuclear}, we obtain $\epsilon_T=-P_{T}(\g sign(\tS) +\epsilon_{\Omega})$. 

Next, following along the lines of Theorem 2 in  \cite{CSPW_siam_2011}, we can show that $\|P_{\Omega^C}(\widehat{\Q})\|_{\infty}<1$ if $\g > \frac{\xi(\tL)}{1-4 \mu(\tS)\xi(\tL)}$ and $\|P_{T^{\perp}}(\widehat{\Q}_1)\|_2<1$ if $\g<\frac{1-3\mu(\tS)\xi(\tL) }{\mu(\tS)}$. The theorem statement follows.

\subsection{Proof of Proposition 5:}
\label{app:PSD_diff_trunc}
The truncation error in estimation of PSD matrix is given by,
\begin{align} 
\nonumber
&\linf{\Phi_{\x}(z)-\widebar{\Phi}_{\x}(z)}\\
&\hspace{1cm}=\Linf{\sum_{k=p+1}^{\infty} \left[R_{\x}(k)z^{-k}+R_{\x}(-k)z^k \right]}\\
&\hspace{1cm}\stackrel{}{\leq} 2\sum_{k=p+1}^{\infty} \linf{R_{\x}(k)} \leq 2C_1 \frac{\rho^{p+1}}{1-\rho},
\end{align} 
where the inequality follows by triangle inequality, since $|z|=1$. The lemma follows by letting $\eps=2C_1 \frac{\rho^{p+1}}{1-\rho}$.

\subsection{Proof of Proposition 6:}

\label{app:linf_PAC_bound}
Define $\Z=[\x_1^T,\dots,\x_N^T]$ and $R=\mathbb{E}\{ \Z\Z^T\}$. Note that $\Z$ is obtained by stacking the observed time series along a column. This structure helps us in bounding deviation of each individual elements.

The following lemmas are useful in the proof of Proposition 6.
\begin{lemma}
The following relation holds true for all matrices $A\in \mathbb{R}^{n \times n}$ \cite{matrix_analysis}:
\begin{align}
\label{eq:inf_spect_norm_equi}
&\frac{1}{n}\|\A\|_2 \leq \linf{\A} \leq \sqrt{n} \|\A\|_2.
\end{align}
\end{lemma}

\begin{lemma}
\label{lem:bruce_symm}
For every symmetric matrix $\S\in \mathbb{R}^{nN \times nN}$ and for every $\eps>0$, the following bound holds for all $N\geq n$
\begin{align}
\nonumber
&\mathbb{P}\left( \frac{1}{N}\Z^T\S\Z > \frac{1}{N}Tr(R\S)+ \eps \right) \\
&\hspace{1cm}\leq \exp\left(-N\min\left\{\frac{\eps^2}{32 \|\S\|_2^2 n^2C_1^2},\frac{\eps}{8 \|\S\|_2 nC_1}\right\} \right).
\end{align}
\end{lemma}

\begin{proof}
The proof is similar to Lemma 4 in \cite{bruce}. The lemma statement follows by plugging in the inequality $\| R\| \leq n\linf{R} \leq nC_1$ from \eqref{eq:inf_spect_norm_equi} and the bound on auto-correlation.
\end{proof}

\begin{lemma}
\label{lem:elementwise_bound}
For every $i,j \in \{1,\dots,n\}$, every $k \in \{0,\dots,p\}$, every $l \in \{1,\dots,N-k\}$, and every ${N-k \geq n}$, the following bound holds:
\begin{align}
\nonumber
&\mathbb{P}\left( \left| [\widehat{R}_{\x}(k)]_{i,j}-[R_{\x}(k)]_{i,j} \right|> \eps \right) \\
&\hspace{1cm}\leq 2\exp\left(-(N-k)\min\left\{\frac{\eps^2}{32  n^2C_1^2},\frac{\eps}{8 nC_1}\right\} \right),
\end{align}
where $[\widehat{R}_{\x}(k)]_{i,j}=\frac{1}{N-k} \sum_{l=1}^{N-k} x_i(l) x_j(l+k)$.
\end{lemma}

\begin{proof}
The idea behind the proof is to pick $\S$ such that $\Z^T\S\Z=\sum_{l=1}^{N-k} x_i(l) x_j(l+k)
$. Designing $\S$ such that $\S_{lm}=1$ if $rem(l,n)=i$ and $(rem(m-nk,n)=j)\& (m-nk>0)$, where $rem(a,b)$ is the remainder of $a/b$ will satisfy this condition. Then, 
\begin{align*}
Tr\{\S R\}&=Tr\{\S\mathbb{E}\{\Z\Z^T\}\}=\mathbb{E}[Tr\{\Z^T\S\Z\}]\\
&=[R_{\x}(k)]_{ij}.
\end{align*}
Lemma \ref{lem:elementwise_bound} follows by applying Lemma \ref{lem:bruce_symm} and repeating the same for -S and for every $i$ and $j$.
\end{proof}
\begin{remark}
Note that in Lemma \ref{lem:elementwise_bound} we have used $N-k$ instead of $N$. The bound still holds, since $N-K\leq N$.
\end{remark}

Now we can prove Proposition 6.

The result is obtained by applying union bound and Lemma \ref{lem:elementwise_bound}.
\begin{align}
\nonumber
&\mathbb{P}\left( \linf{\widehat{R}_{\x}(k)-R_{\x}(k)}> \eps \right) \\
&\hspace{.3cm}\leq  \mathbb{P}\left( \bigcup_{i,j=1}^n \left| [\widehat{R}_{\x}(k)]_{i,j}-[R_{\x}(k)]_{i,j} \right|> \eps \right)\\
&\hspace{.3cm}\leq \sum_{i,j=1}^n \mathbb{P}\left( \left| [\widehat{R}_{\x}(k)]_{i,j}-[R_{\x}(k)]_{i,j} \right|> \eps \right),
\end{align}
which gives the desired result by plugging in Lemma \ref{lem:elementwise_bound}.

\subsection{Proof of Lemma 7:}
\label{app:finite_error_pac}
From Proposition 6, we have 
\begin{align}
\nonumber
&\mathbb{P}\left( \linf{\widehat{R}_{\x}(k)-R_{\x}(k)}\leq \eps \right) \\
&\hspace{.5cm}\geq1- n^2 \exp\left(-(N-k)\min\left\{\frac{\eps^2}{32 n^2C_1^2},\frac{\eps}{8 nC_1}\right\} \right).
\end{align}
Then,	\begin{align}
\sum_{k=-p}^{p} \linf{R_{\x}(k)-\widehat{R}_{\x}(k)} \leq  (2p+1)  \eps
\end{align}
with probability at least \begin{align*}
1- n^2 \exp\left(-(N-p)\min\left\{\frac{\eps^2}{32 n^2C_1^2},\frac{\eps}{8 nC_1}\right\} \right).
		\end{align*}
		Since $\eps$ is arbitrary, the lemma follows by picking appropriate epsilon and applying the inequality (26).

\subsection{Proof of Theorem 7:}
\label{app:IPSD_error_pac}
The following lemma from \cite{matrix_analysis} is useful in deriving this.
\begin{lemma}
\label{lem:matrix_perturbation}
For any invertible matrices $\A$ and $\B$ with $\|\A^{-1}\|_2\|\B-\A\|_2<1$, we have 

{\footnotesize
\begin{align}	\|\A^{-1}-\B^{-1}\|_2 \leq\|\A^{-1}\|_2\|\A\|_2^{-1}\| \B-\A\|_2 \frac{\kappa(\A) }{1-\kappa(\A) \frac{\|\B-\A\|_2}{\|\A\|_2}},
\end{align}}

where $\kappa(\A)$ is the condition number of $\A$.
\end{lemma}
	
Let $\A=\Phi_{\x}(z)$ and $\B=\widehat{\Phi}_{\x}(z)$. Then, by applying Lemma \ref{lem:matrix_perturbation}, we get
\begin{align}
\nonumber
&\|\Phi_{\x}^{-1}-\widehat{\Phi}_{\x}^{-1}\|_2 \\
\label{eq:inverse_ineq}
&\hspace{.5cm} \leq \|\Phi_{\x}^{-1}\|_2\|\Phi_{\x}\|_2^{-1}\| \widehat{\Phi}_{\x}-\Phi_{\x}\|_2 \frac{\kappa(\Phi_{\x}) }{1-\kappa(\Phi_{\x}) \frac{\|\widehat{\Phi}_{\x}-\Phi_{\x}\|_2}{\|\Phi_{\x}\|_2}}.
\end{align}
From the definition (23), and by applying sub-multiplicative property of the spectral norm, $\|\Phi_{\x}^{-1}(z)\|_2$ $\leq \| (\I-H(z))^{*} \|_2 \| \Phi^{-1}_e(z)\|_2 \| (\I-H(z))\|_2$ 
$\leq \frac{1}{l^2 \sigma_{e_{[n_o]}}^2}$, where $\sigma_{e_{[n_o]}}$ is the smallest eigenvalue of $\Phi_e(z)$. Similarly, $\kappa(\Phi_{\x}):=\frac{\lambda_{max}(\Phi_{\x})}{\lambda_{min}(\Phi_{\x})} \leq \frac{L^2 \sigma_{e_{[1]}}^2 }{l^2\sigma_{e_{[n_o]}}^2}.$

Then,
\begin{align}
\nonumber
&\|\Phi_{\x}^{-1}-\widehat{\Phi}_{\x}^{-1}\|_2 \\
\label{eq:Phi_inv_final}
&\hspace{.5cm} \leq \frac{L^2 \sigma_{e_{[1]}}^2}{l^4 \sigma_{e_{[n_o]}}^4}\left(\frac{\| \widehat{\Phi}_{\x}-\Phi_{\x}\|_2 }{l^2 \sigma_{e_{[n_o]}}^2-\|\widehat{\Phi}_{\x}-\Phi_{\x}\|_2}\right).
\end{align}
The theorem follows by applying \eqref{eq:inf_spect_norm_equi}.

\subsection{Proof of Theorem 5:}
\label{appendix:two_Par_Sp}


{\color{black} First, notice that the following holds in (35) based on Lemma $6$. 
\begin{enumerate}
    \item if $a_3\neq 0,$ then $a_3 \in \R:$ the term $a_3$ corresponds to $l \rightarrow k \leftarrow i,$ which implies that $l$ and $i$ are spouses. 
    \item if $b_3\neq 0,$ then $b_3 \in \R:$ the term $b_3$ corresponds to $l \rightarrow k \leftarrow j,$ which implies that $l$ and $j$ are spouses. 
    \item if $a_2b_2 \neq 0,$ then $a_2b_2 \in \R:$ the term $a_2b_2$ corresponds to $i\rightarrow l \leftarrow j,$ which implies that $i$ and $j$ are spouses. By similar argument, $\Im\{[\H_{ho}(l,i)]^{*}\Phi_{e_{h,l}}^{-1}\H_{ho}(l,j)\}=0$
\end{enumerate}
Hence, $\Im(a_2b_2), \Im(a_3)$ and $ \Im(b_3)$ are zero. Then, 
\small
\begin{align}\label{Imag_L}
&{\Im (\L)}_{ij} =  - \sum_{l=1}^{n_h} d_l[\Im(a_1b_1)+\Im(a_1b_2)+\Im(a_1b_3)+\Im(a_2b_1)+\nonumber \\
&\ \ \ \ \ \ \ \ \ \ +\Im(a_2b_3)+ \Im(a_3b_1)+\Im(a_3b_2)]
\end{align}
\normalsize

\begin{remark}
\label{rem:proof_rem}
Under Assumption 3, 4, for $i,j \in M_l$, such that  (a) $i,~j$ are not strict parents or (b) $i,~j$ are not strict spouses of hidden node in $\mV_h$, then from \eqref{Imag_L} it follows that $\Im\{\L_{ij}\} \neq 0$ almost always.
\end{remark}

{\color{black} \begin{remark}
Under Assumption 4, we apply Theorems 2 and 3, but by replacing $\L_{ij}$ in the definitions with $\Im\{\L_{ij}\}$ and $DM_h(i,j)$ with $DE_h(i,j)$. In other words, if $\Im\{\L_{ij}\}$ is not equal to zero, then there exists a $g \in DE_h(i,j)$ in $\mG(\mV,\mE)$.
\end{remark}
}

}
$(\Rightarrow)$ 
From Assumption 2, it implies that $\mathcal{C}(l) \cup \mathcal{P}(l) \cup \mathcal{S}(l) \in \mV_o.$ We use contrapositive argument to show this, i.e., we will show that if $|\mathcal{P}(l)\setminus\{\mathcal{C}(l)\cup \mathcal{S}(l) \}|\leq 1$ and $|\mathcal{S}(l)\setminus\{\mathcal{C}(l)\cup \mathcal{P}(l) \}|\leq 1,$ then $\nexists k \in M_l$ such that $deg_{M_l}(k)< \alpha_l.$ Here, we will show that for any node $a \in M_l$, $deg_{M_l}(a)=|M_l|-1$.

\noindent Case $1$: \footnotesize
$|\mathcal{P}(l)\setminus(\mathcal{C}(l)\cup \mathcal{S}(l))|=0, |\mathcal{S}(l)\setminus\{\mathcal{C}(l)\cup \mathcal{P}(l) \}|=0:$
\normalsize

Here, hidden node $l$ does not have either strict parent and strict spouse, which implies that $M_l = \mathcal{C}(l) \cup (\mathcal{P}(l)\cap \mathcal{S}(l)).$ 
Consider an observable node $a_1 \in \mathcal{C}(l) \cup (\mathcal{P}(l)\cap \mathcal{S}(l));$ existence of $a_1$ is guaranteed from Assumption 1. Suppose there exists $a_2 \in \mathcal{C}(l) \cup (\mathcal{P}(l)\cap \mathcal{S}(l)),\  a_2 \neq a_1$. If $a_1 \in \mathcal{C}(l),$ then clearly there exists a $g \in DE_l(a_1,a_2)$ such that $ g \in \mG(\mV,\mE)$. Suppose $a_1 \in (\mathcal{P}(l)\cap \mathcal{S}(l))\setminus \mathcal{C}(l)$. Then, again, from the definition of $DE_l(a_1,a_2)$,  $\exists g \in DE_l(a_1,a_2)$ such that $ g \in \mG(\mV,\mE)$. It follows from {\color{black}Remark \ref{rem:proof_rem}} that ${\Im\{\L\}}_{a_1a_2}\neq 0$ and ${\Im\{\L\}}_{a_2a_1}\neq 0$ almost everywhere, which implies $(a_1,a_2)\in Q_l$. Essentially, we have shown that for every $a_1 \in M_l$ and for every $k \in M_l\setminus\{a_1\}$, $(a_1,k)\in Q_l$, i.e., $deg_{M_l}(a_1)=|M_l\setminus\{a_1\}|= |M_l|-1.$ The maximum degree is $\alpha_l =|M_l| -1,$ and thus, $\nexists k \in M_l$ such that $deg_{M_l}(k)< \alpha_l.$ 
We use the following conclusion in proof of Case $2,3$ and $4$: Consider a node $a_1\in \mathcal{C}(l)\cup (\mathcal{P}(l)\cap \mathcal{S}(l)).$ Then, for every $a_2 \in \mathcal{C}(l)\cup (\mathcal{P}(l)\cap \mathcal{S}(l))\setminus\{a_1\}, (a_1,a_2)\in Q_l.$ 

\noindent Case 2: {\footnotesize
$|\mathcal{P}(l)\setminus\{\mathcal{C}(l)\cup \mathcal{S}(l)\}|=0,|\mathcal{S}(l)\setminus\{\mathcal{C}(l)\cup \mathcal{P}(l) \}|=1:$}

Here, there is no strict parent. One strict spouse $s_1$ is present, which implies that $M_l = \mathcal{C}(l) \cup (\mathcal{P}(l)\cap \mathcal{S}(l))\cup s_1.$ 
Consider any observable node $a_1 \in \mathcal{C}(l) \cup (\mathcal{P}(l)\cap \mathcal{S}(l)).$ Existence of $a_1$ is guaranteed by Assumption 1. It follows from Case $1$ that $(a_1,a_2)\in Q_l, \forall a_2 \in \mathcal{C}(l)\cup (\mathcal{P}(l)\cap \mathcal{S}(l))\setminus{a_1}.$ We next show that $(a_1,s_1) \in Q_l$. Since $s_1$ is strict spouse, there exists a $g \in DE_l(a_1,s_1)$  such that $ g \in \mG(\mV,\mE).$ Then, by {\color{black}Remark \ref{rem:proof_rem}}, ${\Im\{\L\}}_{a_1s_1}\neq 0$ and ${\Im\{\L\}}_{s_1a_1}\neq 0$ almost everywhere, which implies that $(a_1,s_1)\in Q_l$. Since $M_l = \mathcal{C}(l) \cup (\mathcal{P}(l)\cap \mathcal{S}(l))\cup s_1,$ we can conclude that for any $a_1 \in M_l\setminus{s_1}, (a_1,s_1)\in Q_l.$ Moreover, $deg_{M_l}(a_1)=|M_l|-1$ and $deg_{M_l}(s_1)=|M_l|-1.$ Note that, here, $\alpha_l=|M_l|-1$.
Thus, $\nexists k \in M_l$ such that $deg_{M_l}(k)< \alpha_l.$
We use the following conclusion in Case $4:$ Consider a node $s_1\in \mathcal{S}(l)\setminus (\mathcal{C}(l)\cup \mathcal{P}(l) ),$ then for any $a_1 \in \mathcal{C}(l)\cup (\mathcal{P}(l)\cap \mathcal{S}(l)), (s_1,a_1)\in Q_l.$

\footnotesize
Case 3: $|\mathcal{P}(l)\setminus\{\mathcal{C}(l)\cup \mathcal{S}(l)\}|=1,|\mathcal{S}(l)\setminus\{\mathcal{C}(l)\cup \mathcal{P}(l) \}|=0:$
\normalsize
Here, there is no strict spouse but one strict parent $p_1$ is present, which implies that $M_l = \mathcal{C}(l) \cup \{\mathcal{P}(l)\cap \mathcal{S}(l)\}\cup p_1.$ Consider any observable node $a_1 \in \mathcal{C}(l) \cup (\mathcal{P}(l)\cap \mathcal{S}(l)).$ Existence of $a_1$ is guaranteed from the Assumption 1. From the conclusion of Case $1$, it follows that for all $a_2 \in \mathcal{C}(l)\cup (\mathcal{P}(l)\cap \mathcal{S}(l))\setminus\{a_1\},$ $(a_1,a_2)\in Q_l.$ We next show that $a_1$ is also connected to $p_1.$ Since $p_1$ is strict parent, there exists a $g \in DE_l(a_1,p_1)$ such that $ g \in \mG(\mV,\mE).$ By {\color{black}Remark \ref{rem:proof_rem}}, ${\Im\{\L\}}_{a_1p_1}\neq 0$ and ${\Im\{\L\}}_{p_1a_1}\neq 0$, which implies that $(a_1,p_1) \in Q_l$. Since $M_l = \mathcal{C}(l) \cup (\mathcal{P}(l)\cap \mathcal{S}(l))\cup p_1,$ we can conclude that $deg_{M_l}(a_1)=|M_l|-1.$ Moreover, for any $a_1 \in M_l\setminus{p_1}, (a_1,p_1)\in Q_l.$ Therefore, $deg_{M_l}(p_1)=|M_l|-1.$ Here, $\alpha_l=|M_l|-1$.
Thus, $\nexists k \in M_l$ such that $deg_{M_l}(k)< \alpha_l.$

\footnotesize
Case $4: |\mathcal{P}(l)\setminus\{\mathcal{C}(l)\cup \mathcal{S}(l)\}|=1,|\mathcal{S}(l)\setminus\{\mathcal{C}(l)\cup \mathcal{P}(l) \}|=1:$
\normalsize
Here, there is one strict parent $p_1$, one strict spouse $s_1,$ 
which implies that $M_l = \mathcal{C}(l) \cup \{\mathcal{P}(l)\cap \mathcal{S}(l)\}\cup \{p_1,s_1\}.$ Consider $a_1 \in \mathcal{C}(l) \cup \{\mathcal{P}(l)\cap \mathcal{S}(l)\}.$ It follows from conclusion of Case $2$ and $3$ that $(a_1,s_1) \in Q_l$ and $(a_1,p_1) \in Q_l$. Thus, $deg_{M_l}(a_1)=|M_l|-1$. Moreover, $\exists g \in DE_l(p_1,s_1)$ such that $g \in \mG(\mV,\mE)$. It follows that $(p_1,s_1) \in Q_l$ and thus $deg_{M_l}(p_1)=deg_{M_l}(s_1)=|M_l|-1$. Here, $\alpha_l=|M_l|-1$ and hence, $\nexists k \in M_l$ such that $deg_{M_l}(k)< \alpha_l.$
This concludes the proof.


$(\Leftarrow)$
For the converse, we have from  Assumption 1 that there exists an $i \in \mathcal{C}(l)$. Then, it follows that $deg_{M_l}(i)=|M_l|-1$. Thus, it suffices to show in each of the following cases that there exists a node $i \in M_l$ with $deg_{M_l}(i)<|M_l|-1$.

Suppose that $|\mathcal{P}(l)\setminus(\mathcal{C}(l)\cup \mathcal{S}(l) )|\geq 2$. Then, there exist distinct $p_1,p_2 \in \mathcal{P}(l)\setminus(\mathcal{C}(l)\cup \mathcal{S}(l))$. Since $p_1$ and $p_2$ are strict parents in $\mG(\mV,\mE)$, there does not exist $g \in DE_h(p_1,p_2)$ for any $h \in \mV_h$ such that $g \in \mG{(\mV,\mE)}$. Hence, by {\color{black}Remark \ref{rem:proof_rem}}, $[\Im\{ \L\}]_{p_1p_2}=0$, and $(p_1,p_2) \notin {Q}_l$. Thus, $deg_{M_l}(p_1)<|M_l|-1$ and $deg_{M_l}(p_2)<|M_l|-1$. 

The similar proof holds if $|\mathcal{S}(l)\setminus(\mathcal{C}(l)\cup \mathcal{P}(l) )|\geq 2$. Suppose that $|\mathcal{S}(l)\setminus(\mathcal{C}(l)\cup \mathcal{P}(l) )|\geq 2$. Then, there exist distinct $s_1,s_2 \in \mathcal{S}(l)\setminus(\mathcal{C}(l)\cup \mathcal{P}(l))$. Since $s_1$ and $s_2$ are strict spouses, there does not exist $g \in DE_h(s_1,s_2)$ for any $h \in \mV_h$ such that $g \in \mG{(\mV,\mE)}$. Hence, by {\color{black}Remark \ref{rem:proof_rem}}, $[\Im\{ \L\}]_{s_1s_2}=0$, and $(s_1,s_2) \notin {Q}_l$. Thus, $deg_{M_l}(s_1)<|M_l|-1$ and $deg_{M_l}(s_2)<|M_l|-1$, which concludes the proof.		

\subsection{LDG example:}
\begin{figure}
\includegraphics[width=.99\columnwidth]{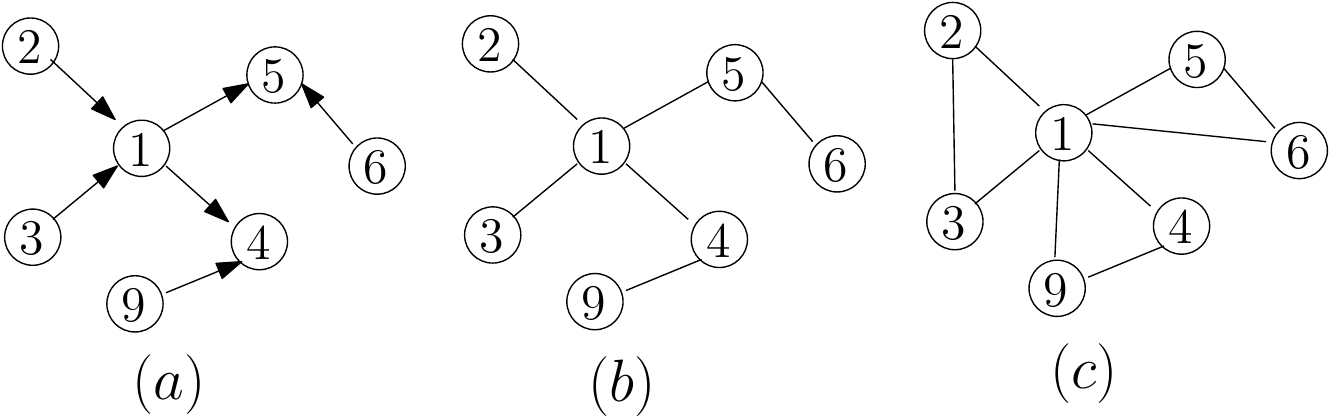}
\caption{(a) LDG $\mG(\mV,\mE),$ (b) $top(\mG)$ and (c) $kin(\mG)$. $\{2,3 \}$ are strict parents of $1.$ $\{6,9 \}$ are strict spouses of $1.$  }
\label{fig:example}
\end{figure}

An example illustrating a LDG, its topology, moral graph, and the definitions of strict parents and strict spouses, is shown in Fig. \ref{fig:example}.
